\documentclass[%
 aip,
 amsmath,amssymb,
 reprint,%
]{revtex4-1}
\usepackage{graphicx}
\usepackage{dcolumn}
\usepackage{bm}
\usepackage[utf8]{inputenc}
\usepackage[T1]{fontenc}
\usepackage{mathptmx}
\usepackage{hyperref}

\usepackage{gensymb}
\usepackage{amsmath}
\usepackage{breqn}
\usepackage{float}
\usepackage[section]{placeins}

\usepackage{scalerel}
\usepackage{tikz}
\usetikzlibrary{svg.path}
\definecolor{orcidlogocol}{HTML}{A6CE39}
\tikzset{
  orcidlogo/.pic={
    \fill[orcidlogocol] svg{M256,128c0,70.7-57.3,128-128,128C57.3,256,0,198.7,0,128C0,57.3,57.3,0,128,0C198.7,0,256,57.3,256,128z};
    \fill[white] svg{M86.3,186.2H70.9V79.1h15.4v48.4V186.2z}
                 svg{M108.9,79.1h41.6c39.6,0,57,28.3,57,53.6c0,27.5-21.5,53.6-56.8,53.6h-41.8V79.1z M124.3,172.4h24.5c34.9,0,42.9-26.5,42.9-39.7c0-21.5-13.7-39.7-43.7-39.7h-23.7V172.4z}
                 svg{M88.7,56.8c0,5.5-4.5,10.1-10.1,10.1c-5.6,0-10.1-4.6-10.1-10.1c0-5.6,4.5-10.1,10.1-10.1C84.2,46.7,88.7,51.3,88.7,56.8z};
  }
}
\newcommand\orcidicon[1]{\href{https://orcid.org/#1}{\mbox{\scalerel*{
\begin{tikzpicture}[yscale=-1,transform shape]
\pic{orcidlogo};
\end{tikzpicture}
}{|}}}}

\makeatletter
\def\@email#1#2{%
 \endgroup
 \patchcmd{\titleblock@produce}
  {\frontmatter@RRAPformat}
  {\frontmatter@RRAPformat{\produce@RRAP{*#1\href{mailto:#2}{#2}}}\frontmatter@RRAPformat}
  {}{}
}%
\makeatother

\begin{document}

\preprint{AIP/123-QED}

\title[MKIDGen3]{MKIDGen3:\\Energy-Resolving, Single-Photon-Counting MKID Readout on an RFSoC}
\author{Jennifer Pearl Smith \orcidicon{0000-0002-0849-5867}}
 \email{jennifer\_smith@ucsb.edu}
 \altaffiliation[]{These authors contributed equally}
\author{John I. Bailey, III. \orcidicon{0000-0002-4272-263X}}%
\altaffiliation[]{These authors contributed equally}
  \author{Aled Cuda \orcidicon{0009-0005-5550-7620}}%
  \author{Nicholas Zobrist \orcidicon{0000-0003-3146-7263}}
 \author{\\Benjamin A. Mazin \orcidicon{0000-0003-0526-1114}}%
 \affiliation{ 
Department of Physics; University of California; Santa Barbara; California 93106; USA
}%

\date{\today}

\begin{abstract}
Microwave Kinetic Inductance Detectors (MKIDs) are superconducting detectors capable of single-photon counting with energy resolution across the ultraviolet, optical, and infrared (UVOIR) spectrum with microsecond timing precision. MKIDs are also multiplexable, providing a feasible way to create large-format, cryogenic arrays for sensitive imaging applications in biology, astronomy, and quantum information. Building large, cryogenic MKID arrays requires processing highly-multiplexed, wideband readout signals in real time; a task that has previously required large, heavy, and power-intensive custom electronics. In this work, we present the third-generation UVOIR MKID readout system (Gen3) which is capable of reading out twice as many detectors with a fifth the weight and power and an order of magnitude less volume and cost-per-pixel as compared to the previous system. Gen3 leverages the Xilinx RFSoC4x2 platform to read out 2048, 1 MHz MKID channels per board. The system takes a modern approach to FPGA design using Vitis High-Level Synthesis (HLS) to specify signal processing blocks in C/C++, Vivado ML Intelligent Design Runs (IDR) to inform implementation stragety and close timing, and Python Productivity for ZYNQ (PYNQ) to simplify interacting with and programming the FPGA using Python. This design suite and tool flow allows general users to contribute to and maintain the design and positions Gen3 to rapidly migrate to future platforms as they become available. In this work, we describe the system requirements, design, and implementation. We also provide performance characterization details and show that the system achieves detector-limited resolving power in the case of few readout tones and minimal degradation with all 2048 tones. Planned upgrades and future work are also discussed. The Gen3 MKID readout system is fully open-source and is expected to facilitate future array scaling to megapixel-sized formats and increase the feasibility of deploying UVOIR MKIDs in space.

\end{abstract}

\maketitle


\section{\label{sec:intro}Introduction}
Microwave Kinetic Inductance Detectors (MKIDs) are promising superconducting detectors with applications for highly-sensitive photon measurements such as those needed in quantum computing, biological imaging, and astrophysical observations \cite{todaro_state_2021,mariantoni_implementing_2011,xia_short-wave_2021,niwa_few-photon_2017,zmuidzinas_superconducting_2012,mazin_microwave_2009}. MKIDs are able to detect the energy and arrival time of single photons using the changes induced in the kinetic inductance of superconducting materials by incident photons \cite{day_broadband_2003}. This operating principle allows for single-photon counting with zero read noise or dark current across ultraviolet, optical, and infrared (UVOIR) wavelengths with microsecond timing precision\cite{zobrist_wide-band_2019,zobrist_membraneless_2022}. MKIDs also natively support multiplexing many detectors per readout line using frequency-division multiplexing (FDM)--a technique that will be discussed more in Sec. \ref{sec:fdm}. FDM provides a feasible means of scaling MKIDs to large cryogenic arrays suitable for high-resolution imaging. 

The focus of our group is to develop UVOIR MKIDs into fast, energy-resolving, single-photon-counting cameras for scientific imaging and spectroscopy. We have successfully deployed a twenty-kilopixel UVOIR MKID array in the MKID Exoplanet Camera (MEC) \cite{walter_mkid_2020} and have demonstrated several high-contrast imaging astronomy results \cite{ steiger_scexaomec_2021, swimmer_scexao_2022}. We are developing an echelle spectrograph testbench \cite{iii_m-most_2022} and are also pursing science applications in quantum information and biological imaging. 

Progress has been slowed in part by the current, second-generation MKID digital readout system (hereafter Gen2) which is excessively large, cumbersome, and power-hungry by modern standards \cite{fruitwala_second_2020}. Gen2 will not scale for larger ground-based arrays and is not suitable for future space-based missions. Gen2 also relies on obsolete hardware and tools, making it exceedingly difficult to update and to integrate modern advances in photon signal detection. 

To alleviate these challenges, we have created the next generation MKID digital readout which is capable of reading out twice as many MKIDs per board with a fraction of the weight, volume, and power of the previous system. A key advancement is the migration to the Xilinx RFSoC platform with FPGA-integrated, high-speed analog-to-digital and digital-to-analog converters (ADCs/DACs). The integrated platform provides a dramatic reduction in power and device footprint that will enable future high-altitude missions and ease system scaling. We also leveraged modern FPGA programming tools including Vitis High-Level Synthesis (HLS), which synthesizes C++ code to FPGA-specific hardware description language, and Python Productivity for Zynq (PYNQ), which facilitates interacting with the FPGA through Python. These tools are more accessible to scientists and astronomers without specialized knowledge of FPGA design and create a system which is easier to maintain and upgrade. 

In this work, we discuss the system design, implementation, and performance. We include a brief introduction to MKID readout in Sec. \ref{sec:mkidreadoutexp} to provide context for the requirements outlined in Sec. \ref{sec:req}. Next, we describe our approach to system design in Sec. \ref{sec:overview} and describe the FPGA implementation including our use of HLS and our timing closure strategy. We provide performance characterization for the system in loopback in Sec. \ref{sec:perf} and include detailed studies of actual cryogenic MKID readout performance in Sec. \ref{sec:mkidreadout}. We conclude with a discussion of the system performance in Sec. \ref{sec:discussion} and describe future directions and next steps. The MKID digital readout described here is fully open-source and available on GitHub under a GPLv3.0 license\footnote{\url{https://github.com/MazinLab/MKIDGen3}}.

\section{\label{sec:mkidreadoutexp} MKID Readout}
\begin{figure}  
  \begin{center}
  \includegraphics[width=\columnwidth]{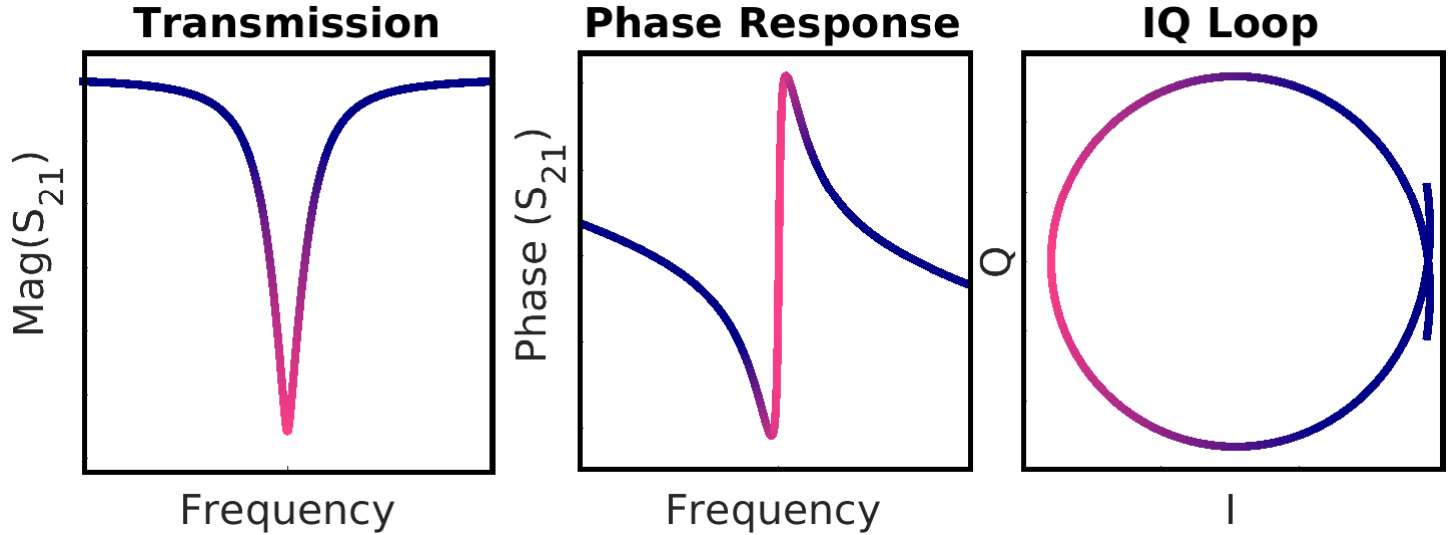}
  \end{center}
  \caption{Simulated MKID frequency response magnitude (left), phase (center), and complex representation (right). The difference between the drive frequency and resonance frequency is shown in a pink to purple gradient on each plot. The optimal readout frequency is near the resonance frequency (pink).}
  \label{fig:loop}
\end{figure}
MKID readout is a rich topic with a large body of prior work dedicated to the underlying physics and improving detector performance \cite{gao_jiansong_physics_2008, mazin_benjamin_a_microwave_2005, zobrist_nicholas_improving_2022}. Here we provide a brief review of core MKID readout signals and the strategy underpinning the digital design. We include a high-level description of MKID setup and operation to clarify goals, terminology, and analysis used in the remainder of this work.

\subsection{MKIDs as Superconducting Resonators}
Each MKID pixel is a superconducting LC resonator. Our designs target resonance frequencies in 4-8 GHz due to commercial availability of cryogenic low-noise amplifiers in this band. MKIDs are read out using a homodyne scheme where the signal of interest is a modulation on the readout tone and can be represented as a complex signal. This signal is acquired by down-converting the readout tone to 0 Hz. A simulated MKID frequency response is plotted in Fig. \ref{fig:loop}. The IQ loop (Fig. \ref{fig:loop}, right) contains all the information needed to determine the correct readout frequency, phase offset, and loop center coordinates for photon readout, and as such characterizing each MKID IQ loop is an important setup function for MKID readout systems.

\subsection{Frequency-Division Multiplexing\label{sec:fdm}}
\begin{figure}  
  \begin{center}
  \includegraphics[width=\columnwidth]{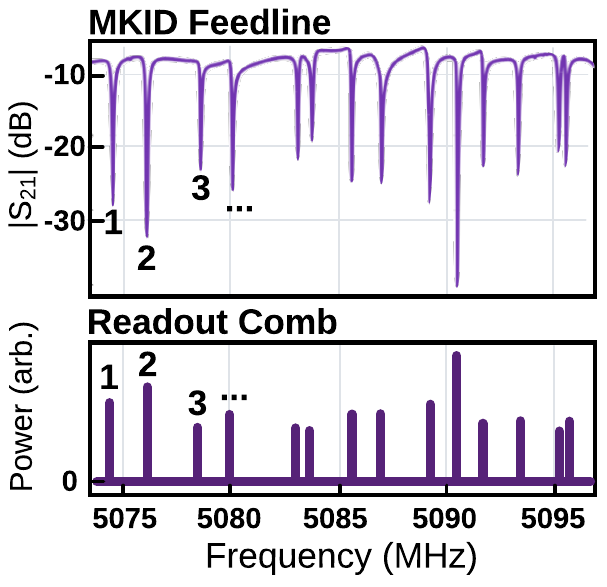}
  \end{center}
  \caption{Overview of readout strategy for a frequency-division-multiplexed MKID feedline. Top: MKID feedline transmission. Each dip in transmission is a different
MKID resonator. Bottom: Cartoon representation of corresponding DAC waveform in Fourier space that would be used to read out this feedline. Every MKID has a corresponding tone in the readout comb near its resonance frequency. The first three MKIDs shown in the feedline and their corresponding readout tones in the frequency comb are numbered.}
  \label{fig:comb}
\end{figure}
MKIDs use frequency-division multiplexing to thread many superconducting resonators on a single readout line. This technique allows MKID readout signals to share a microwave feedline, conserving cryogenic heat load and promoting the creation of large arrays. A sample transmission from a real MKID feedline is shown in Fig. \ref{fig:comb}, (top). Each resonator is fabricated with a different resonance frequency, allowing each resonator to be addressed by its unique resonance. We target 2048 resonators with a spacing of 2 MHz but lose 10\% to 40\% of pixels due to resonator collisions and other fabrication process uncertainties. More details on MKID array fabrication are provided in references\cite{szypryt_high_2016, coiffard_characterization_2020, mazin_superconducting_2020}. Each feedline represents a base unit for a full MKID array with each feedline being an imperfect copy of the others. In this work, references to multi-MKID readout refer to a single feedline with the understanding that a high-level control program can aggregate data from each feedline to create array-level data products.

\subsection{MKID Feedline Setup}
After fabrication, the precise locations of the pixels and their optimal readout parameters are unknown. The digital readout uses a reprogrammable local oscillator (LO) to sweep a comb of uniform tones across the 4-8 GHz feedline while collecting transmission data, similar to a vector network analyzer. The frequency sweep is repeated at different powers and the resulting IQ loops are used to determine the optimal readout frequency, power, phase offset, and loop center coordinates for each MKID. The readout powers and frequencies are used to generate a customized DAC output consisting of a superposition of the readout tones (see Fig. \ref{fig:comb}, bottom). The phase offsets and loop center coordinates are used to rotate and center each IQ loop in order to standardize the photon response curve. After the optimal readout tones and powers have been used to generate the readout waveform and the IQ loops have been rotated and centered, the device is ready for photon readout.

\subsection{Photon Readout}
\begin{figure}  
  \begin{center}
  \includegraphics[width=\columnwidth]{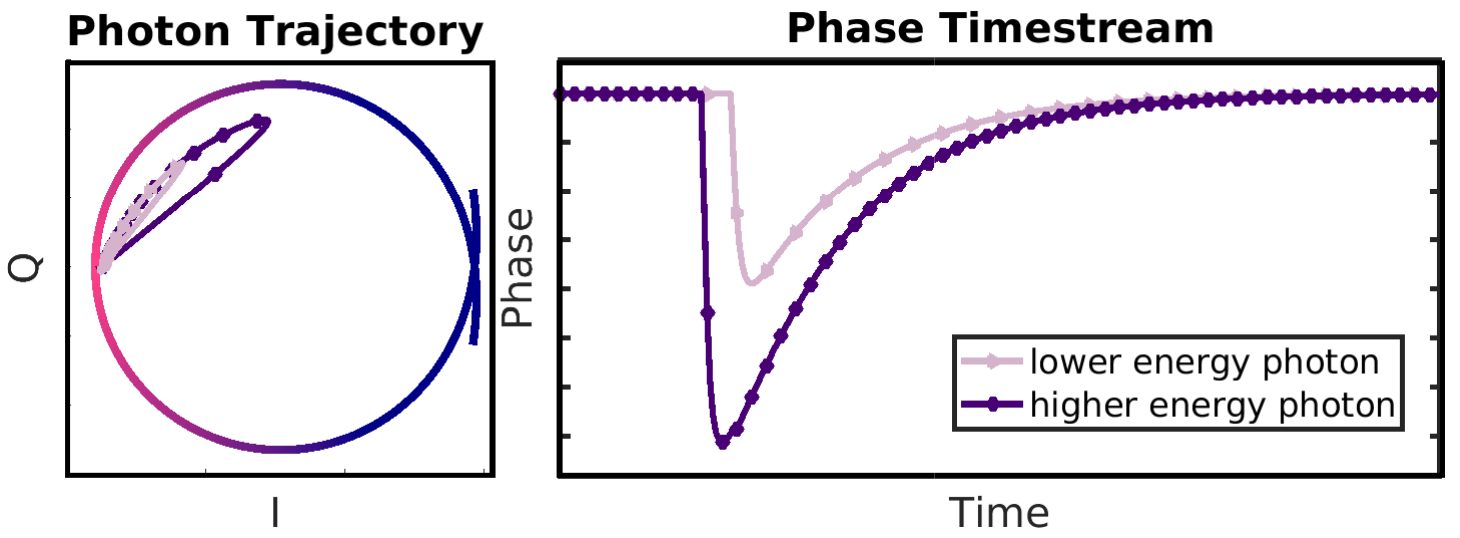}
  \end{center}
  \caption{Simulation of MKID photon readout. Left: the MKID resonator loop is shown in a pink to purple gradient representing the frequency difference from the resonant frequency. The light purple and dark purple lines show the trajectory the resonant frequency takes when a low energy and high energy photon hits the MKID, respectively. Right: The phase of the resonant frequency versus time for the two photon impacts. The depth of the phase pulse corresponds to the energy of the photon. The two photons were constructed to hit at different times for visual clarity.}
  \label{fig:mkidphoton}
\end{figure}

When a photon strikes the detector, the down-converted readout tone follows a trajectory through the IQ plane simulated in Fig. \ref{fig:mkidphoton}, left. To simplify readout, we compute the phase of the trajectory, $\mathrm{tan}^{-1}(Q/I)$, and use this one-dimensional signal to characterize the incident photon (see Fig. \ref{fig:mkidphoton}, right). The start of the phase pulse signifies when the photon hit the detector and the pulse height corresponds to the photon energy.

\subsection{Resolving Power\label{sec:resolvingpowerexp}}
Our ability to determine the energy of the incident photon is an important science metric.
The resolving power, $R$, is defined as
\begin{equation}
R = \frac{E}{\Delta E} \simeq \frac{\lambda}{\Delta \lambda}, \label{r}
\end{equation}
where $E$ is the photon energy and $\Delta E$ is the energy uncertainty. The relation can be expressed in wavelength-space assuming $\Delta E$ is small, but in this work we do all calculations and analysis in energy-space. $R$ is fundamentally limited by the detector design but can also be negatively impacted by phase noise in the probe tone from readout electronics.

$R$ is characterized using a series of lasers with known wavelength. Each laser produces counts with varying pulse heights corresponding to a single photon energy. The full-width-half-max of the distribution of pulse heights is the energy uncertainty. The measured mean pulse height is fit to known laser energy to produce a mapping from phase to energy. Full details of $R$ characterization are available in reference\cite{zobrist_nicholas_improving_2022}.

\section{\label{sec:req}Requirements}
The next generation MKID digital readout system must perform several key functions and adhere to science-driven noise requirements while meeting hardware constraints. The overarching goal is to serve as the digital readout system for a 20,000+  pixel UVOIR MKID instrument. The system is expected to both set up and read out the MKID array. Key objectives, requirements, performance criteria, and constraints are explored in the following sections.

\subsection{Key Objectives}
Several objectives and philosophies played into the system design. Most importantly, we wanted to maximize the number of detectors read out while minimizing the weight, volume, and power of the readout electronics. We also wanted to create a system that is straightforward to migrate to future platforms. This will allow us to continually take advantage of advances in analog-integrated electronics driven by commercial industry.

Another goal was to create a system that is easy for scientists and others without prior FPGA design experience to use, maintain, and improve. In service of this goal and to maximize future collaboration and scientific utility, the system design favors open-source tools and uses more-approachable high-level-synthesis wherever possible.

\subsection{Functional Requirements}
There are several functions the system must perform to effectively read out an MKID feedline. To begin the setup and readout process, the system must be able to generate a 4-8 GHz readout waveform with up to 2048 run-time-programmable frequencies and powers. Each of the 2048 user-defined MKID readout frequencies constitutes a readout channel. For every independent channel, the system must be able to do the following continuously and in real-time:
\begin{itemize}
    \item Down-convert the channel.
    \item Apply a translational and rotational coordinate transform to the down-converted signal.
    \item Calculate the channel phase.
    \item Apply a custom, user-defined filter to the channel phase.
    \item Continuously monitor the channel phase, searching for and recording photon events: the arrival time, pulse height, and channel number.
\end{itemize}
For setup and calibration, the system must be able to capture the following data in finite intervals:
\begin{itemize}
    \item The full data rate input waveform.
    \item The IQ loop of every channel.
    \item The phase time stream of every channel.
\end{itemize}
These captures must each be able to occur simultaneously with photon monitoring as they can provide valuable debug information during an observing run. To debug photon triggering, the system must also be able to simultaneously record a subset of user-defined IQ and phase time streams for a short interval around each photon trigger.

\subsection{\label{sec:snrreq}Signal and Noise Requirements}
Signal and noise requirements for UVOIR MKID readout can be complex, often involving application-specific trade-offs. In this section, we list the the main performance criteria that are common to most use cases and drive design decisions. Requirements are summarized in Table \ref{tab:req}.

\begin{table*}
\caption{\label{tab:req}Summary of signal and noise performance requirements and achievements.}
\begin{ruledtabular}
\begin{tabular}{lllll}
\textbf{Signal or Noise Metric} & \textbf{Requirement} & \textbf{Requirement Section} & \textbf{Achieved} & \textbf{Performance Section}\\
\hline
Readout Tone Frequency Resolution & $\le$ 7.813 kHz & Section \ref{sec:tonegen} & 7.813 kHz & Section \ref{sec:dacreplay}\\
Readout Tone Amplitude Control  & $\le$ 1 dB & Section \ref{sec:tonegen} & 0.25 dB & Section \ref{sec:if}\\
Channel Cross Talk & $\le$ -30 dB & Section \ref{sec:ctimd} & $\le$ -30 dB\footnote{Achieved in RFDC loopback. See Section \ref{sec:discussion} for full discussion.} & Section \ref{sec:combperf}\\
Intermodulation Spurs & $\le$ -30 dB & Section \ref{sec:ctimd} &$\le$ -30 dB$^{\mathrm{a}}$ & Section \ref{sec:combperf}\\ 
Resolving Power, Single MKID & $R\ge7$ at 808 nm & Section \ref{sec:rp} & $R=7$ at 808 nm\footnote{Detector limited. See Sec. \ref{sec:discussion} for full discussion.} & Section \ref{sec:mkidreadout}\\
Resolving Power, 2048 MKIDs & $R\ge4$ at 808 nm & Section \ref{sec:rp} & $R=4$ at 808 nm & Section \ref{sec:mkidreadout}\\
Timing Resolution & $\le$ 1 $\mu$s & Section \ref{sec:timing} & 1 $\mu$s & Section \ref{sec:timestamps}\\
Absolute Timing Precision\footnote{Max error relative to UTC.} & $\le$ 1 $\mu$s & Section \ref{sec:timing} & $\le$ 500 ns & Section \ref{sec:timestamps}\\
\end{tabular}
\end{ruledtabular}
\end{table*}

\subsubsection{Tone Generation\label{sec:tonegen}}
Readout tone frequency precision will impact the detector's perceived responsively and linearity. Errors in tone amplitude affect responsively as well as the phase noise floor. The magnitude of these effects depends on device characteristics including the resonator quality factor and asymmetry. We demonstrated 7.813 kHz frequency resolution and 1 dB amplitude control is sufficient to avoid noticeable performance degradation in the Gen2 readout\cite{fruitwala_second_2020}.

\subsubsection{Cross Talk and Intermodulation Distortion\label{sec:ctimd}}
Cross talk between channels and spurious signals from intermodulation distortion (IMD) cause line noise in the phase time stream that can distort pulse heights and ultimately degrade resolving power. The worst line noise comes from cross talk in the cryogenic MKID device and images generated by gain and phase imbalance in the IQ analog signal chain. Both noise sources can produce spurious signals 20 to 30 dB down from the read out tones. Presently, the IQ mixers are necessary to interface the cryogenic and room temperature electronics and so we require the digital readout crosstalk and IMD to be no worse than -30 dB as referenced to the readout tones.

\subsubsection{Resolving Power\label{sec:rp}}
The resolving power is an important figure of merit used to characterize the detector's ability to discern the energy of the incident photon (see Sec. \ref{sec:resolvingpowerexp}). Pinning down a resolving power requirement specific to the digital readout can be challenging because many other factors including the detector design and fabrication, the cryogenic system and signal chain, the setup and calibration quality, and the off-line data analysis methods can all impact the measurement. To develop our requirement, we bound our system between two extremes: a best-case scenario and a worst-case scenario.  

The best-case scenario performance should about match our in-lab, analog-based, MKID readout which can only read out a single MKID at a time and is nominally used to provide feedback for detector design and fabrication cycles. This system uses a dilution refrigerator (20 mK) with a first-stage, quantum-noise-limited amplifier (TWPA), commercial analog electronics, and Python analysis package with extensive device modeling to measure the best-possible resolving power of individual MKID devices. \citet{zobrist_wide-band_2019} demonstrated a resolving power of 8.9 at 808 nm on a PtSi  MKID\cite{szypryt_large-format_2017} with this system. However, the analysis methods included a wavelength-specific, 500-point+ matched filter and resonator-specific, two-dimensional, quadratic coordinate transform\cite{zobrist_improving_2021}-- both of which are not feasible for an FPGA-based instrument at our target scale. To compensate for the difference in analysis methods, we relax our best-case scenario requirement to $R\ge7$ when reading out a single PtSi, array-style MKID with 808 nm photons.

The worst-case scenario performance should be better than or equal to what was achieved by the previous Gen2 system in the field. Gen2 was deployed to the summit of Mauna Kea with the MKID Exoplanet Camera (MEC) which features an adiabatic de-magnetization refrigerator (90 mK) with a first-stage, Low Noise Factory HEMT amplifier. The Gen2 Python code uses a machine-learning-based approach to semi-automate the set up and read out of 1024 pixels per board. MEC achieved a median R of 4 at 850 nm across the PiSi MKID array. This sets our worst-case scenario requirement to $R\ge4$ at 808 nm\footnote{We set our requirement at 808 nm despite the fact MEC was characterized at 850 nm due to difference in availability of laser sources between Subaru Telescope and our lab at UCSB.} for a PtSi array-style MKID when all 2048 readout channels are active in Gen3.

\subsubsection{Timing Precision\label{sec:timing}}
Millisecond photon arrival resolution is sufficient for most astronomical sources with the notable exception of pulsar timing studies, which benefit from microsecond precision\cite{strader_excess_2013}. In quantum photonics applications, microsecond or better timing precision is preferred\cite{guo_high-timing-precision_2023}. In UVOIR MKID devices, the pulse relaxation time is typically on the order of microseconds. As demonstrated in the Gen2 system\cite{fruitwala_second_2020}, a sample rate of 1 microsecond is sufficient to resolve the pulse peak to within a microsecond and provides enough bandwidth to facilitate signal processing techniques aimed at improving the resolving power.

\subsection{Constraints}
In addition to meeting the performance goals, the new digital readout must adhere to several practical constraints.

\subsubsection{Cryogenic Interface and MKID Bandwidth}
The UVOIR MKIDs are designed with resonance frequencies in the 4-8 GHz band because this is where cryogenic low-noise amplifiers and other specialized components are commercially available. The readout system must be able to supply, sample, and process signals in this bandwidth.

\subsubsection{Hardware}
Due to budget and time restrictions, we are unable to support designing or commissioning a custom analog or digital board and must use existing technology. There are presently no suitable commercial options for the analog interface between the DAC output and cryogenic input, so we must re-purpose existing Gen2 IF boards\cite{fruitwala_second_2020}. 

\subsubsection{Data Rate}
The system must be able to support a count rate of 5,000 counts per second on every pixel. The system should not drop any registered photon data less than this level and should notify users if photons are being dropped.

\subsubsection{Setup Time and Stability}
In the past, MKID instruments have been deployed with adiabatic demagnetization refrigerator (ADR) cryogenic systems. ADRs are well-suited for observatory environments, but impose a time constraint because they can only keep the MKID array cold for a finite hold time. The readout system must be able to perform all setup and calibration steps within this hold time with enough time left over to conduct science observations. In practice, the readout must be able to set up an array from scratch in a few hours. Similarly, the readout must be stable and able to continuously observe and record photons for a typical observing night (at least eight hours).

\begin{figure}  
  \begin{center}
  \includegraphics[width=\columnwidth]{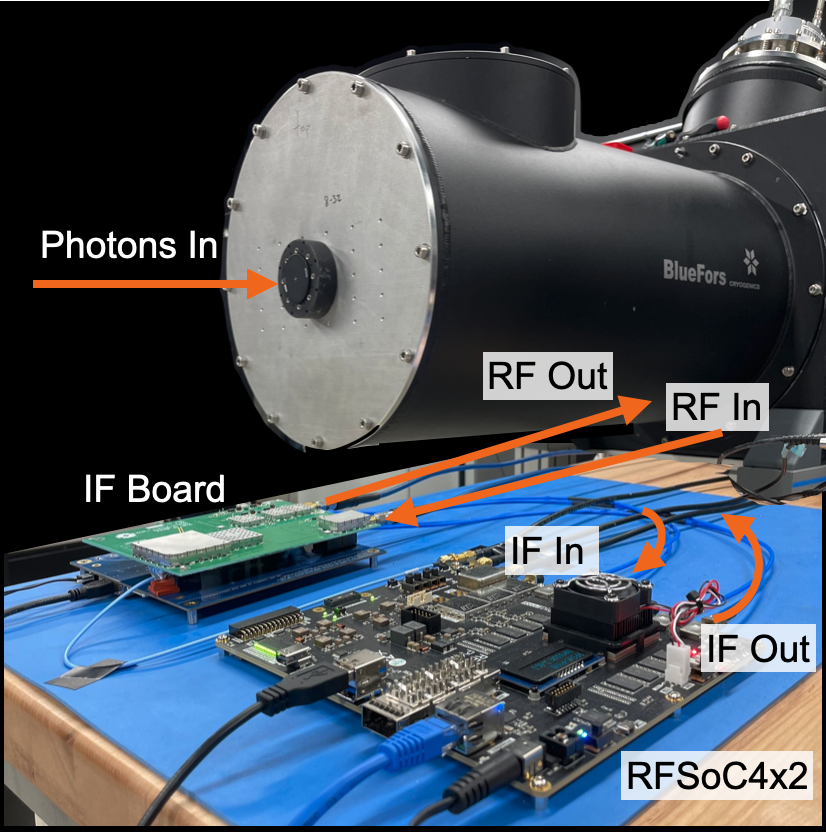}
  \end{center}
  \caption{Photo of MKID digital readout during MKID readout. MKIDs (not shown) are housed at 90 mK inside the dilution refrigerator. Light is passed through the window on the dilution refrigerator front plate. Microwave multiplexed readout signals are driven and acquired by the RFSoC and IF boards via a cryogenic signal chain (not shown). Photon data is saved to network storage over 1 Gigabit Ethernet.}
  \label{fig:experimental_setup_pic}
\end{figure}

\section{\label{sec:overview}System Design}
The Gen3 readout system is based on a Xilinx, analog-integrated FPGA (RFSoC). The integrated platform removes the need for external ADCs and DACs, greatly reducing power consumption and device footprint. At present, the fastest available integrated ADC sampling rate is 5.9 GSPS in the Xilinx RFSoC DFE device\cite{noauthor_zynq_nodate}. To directly sample the 4-8 GHz readout band, the DFE ADC would have to be operated in the third Nyquist zone which is not naively supported\cite{noauthor_rf-adc_nodate}. With direct RF seemingly still out of reach, we opted to build the system around the slower Gen3 RFSoC part which is available in a small, cost-effective academic board: the RFSoC4x2. 

The RFSoC4x2 data converters generate and sample the MKID readout tones using quadrature sampling. An intermediate frequency (IF) mixing board is used to translate the quadrature signals to and from the microwave readout band.
The FPGA programmable logic implements the main digital signal processing pipeline and calibration capture functions. The RFSoC4x2 processing system is used to command and configure the programmable logic as well as peripherals on the IF board. All system communication and data transfer happens over 1 Gigabit Ethernet. 

The readout platform can be run via a Jupyter notebook, acting as a lab-based MKID diagnostic platform, or via a client/server software that facilitates full feedline setup and multi-board photon readout\cite{john_i_bailey_iii_mkidgen3_nodate}. Subsystems are detailed in the following sections.

\begin{table}
\caption{\label{tab:hardware_compare}Gen2 vs. Gen3 system specifications and capabilities.}
\begin{ruledtabular}
\begin{tabular}{lcr}
Specifications\footnote{Per board unless otherwise specified.}&Gen2: ROACH2&Gen3: RFSoC4x2\\
\hline
Dimensions & 24"x12"x4" & 10"x6"x2"\\
Weight & 10 kg & 2 kg \\
RF Bandwidth & 2 GHz & 4 GHz\\
MKID Pixels & 1024 & 2048\\
Data Rate & 40 MiB/s & 80 MiB/s + 16 GiB/s \\
Cost & \$10/pixel & \$3/pixel \\
Power & 175 mW / pixel & 25 mW / pixel\\
Network & 1 GbE & 1 GbE +  100 GbE\footnote{100 GbE is available on the RFSoC4x2 but is not used.}\\
Design System & ISE/Simulink/CASPER & Vivado/PYNQ/HLS \\
Control & Python 2.7 & Python 3.11 / ZeroMQ 
\end{tabular}
\end{ruledtabular}
\end{table}

\begin{figure}  
  \begin{center}
  \includegraphics[width=\columnwidth]{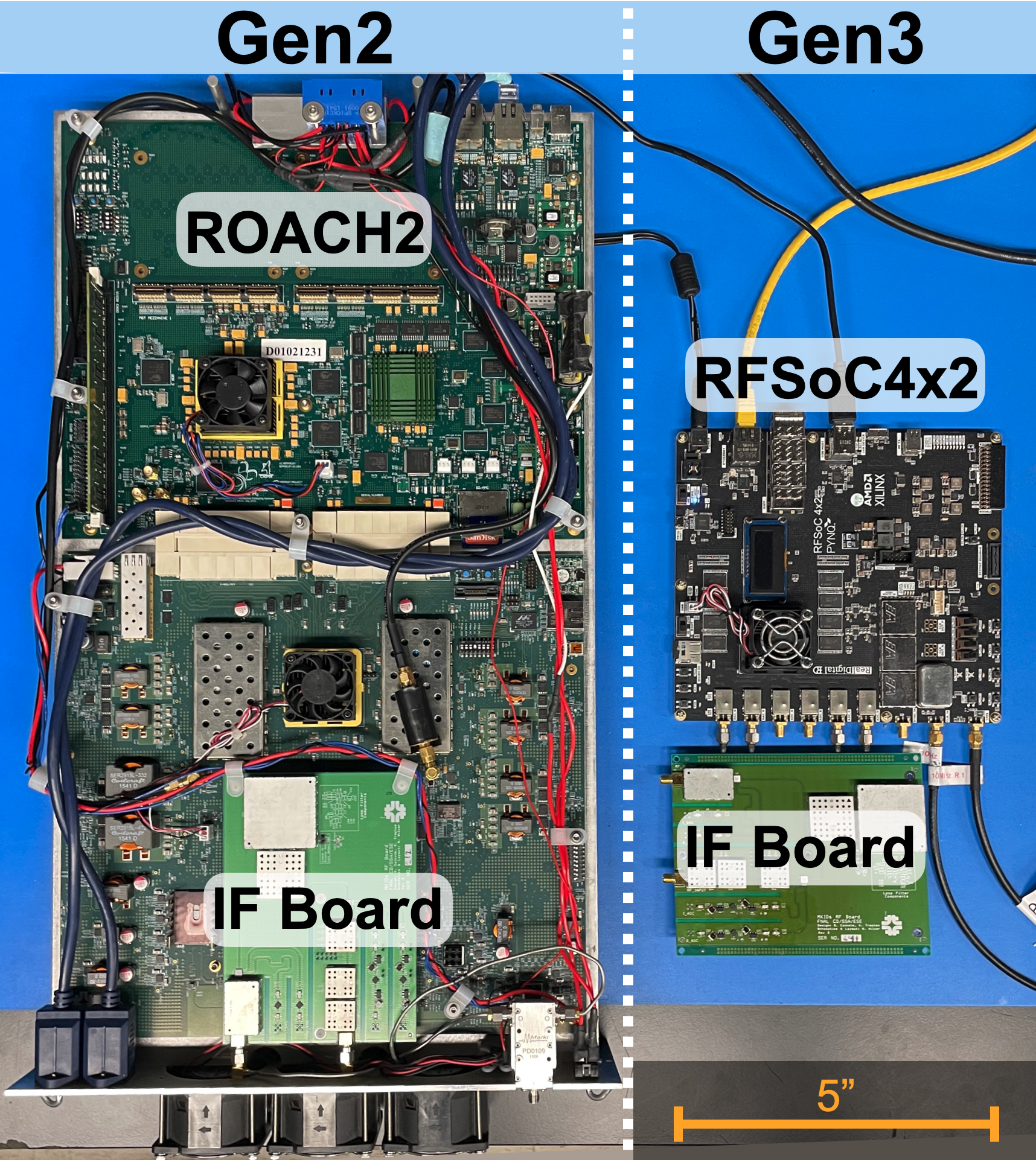}
  \end{center}
  \caption{Scale photo comparison of previous system (Gen2) ROACH2 hardware (left) and Gen3 RFSoC4x2 hardware (right). The same IF board is used in both systems with a few modified components. The Gen3 RFSoC4x2 is significantly smaller, lighter, and more power efficient and is capable of reading out 2x as many detectors as the ROACH2.}
  \label{fig:hardware_comparison_pic}
\end{figure}

\subsection{\label{sec:hardware}Hardware}
A picture of the Gen3 readout hardware and experimental setup is shown in Fig. \ref{fig:experimental_setup_pic}. Migrating to the RFSoC4x2 platform resulted in a dramatic reduction in the weight, volume, and power of the readout electronics. In addition to using only one fifth the power as the previous Gen2 system, the data converters are twice as fast, allowing us to double the number of detectors read out per board. A to-scale picture showing the size reduction from Gen2 to Gen3 is shown in Fig. \ref{fig:hardware_comparison_pic}. Key specifications between the two systems are tabulated in Table \ref{tab:hardware_compare}.

\subsubsection{\label{sec:if} IF Board and Carrier}
The IF board is responsible for translating the $\pm$2 GHz readout waveform to the 4-8 GHz microwave band. It is a modified version of the same board used to convert signals in the Gen2 system \cite{fruitwala_second_2020}. To accommodate the RFSoC's increased bandwidth, we swapped the 630 MHz (LFCN-630+) anti-aliasing filters on the I and Q DAC output \textit{IF to RF} paths for the 2 GHz version (LFCN-2000+). We also swapped 800 MHz (LFCN-800+) and 1800 MHz (LFCN-1800+) anti-aliasing filters on the I and Q \textit{RF to IF} ADC input signal paths with the 2 GHz (LFCN-2000+) and 3.8 GHz (LFCN-3800+) versions, respectively. 

The input and output RF ports each have two programmable attenuators which together can be adjusted from 0 to 63 dB in steps of 0.25 dB. This allows adjusting the waveform power going into the fridge, to optimally drive the MKID resonators, and into the RFSoC, to utilize the full ADC dynamic range. The IF board receiver path has an amplifier chain totaling 88 dB of gain to boost the small cryogenic signals. The board also features a TRF3765 programmable LO and complex mixers which can be used to sweep RFSoC-driven frequencies to different parts of the microwave readout band. 

The IF board is powered by a simple 12 V carrier board (see blue board in Fig. \ref{fig:experimental_setup_pic} under the IF Board). The programmable components are controlled by the carrier's Arduino Nano which is commanded by the RFSoC over USB.

\subsubsection{\label{sec:rfsoc4x2} RFSoC4x2}
Both RFSoC4x2 DACs and two of the ADCs are run at 4.096 GSPS to create a quadrature sampled readout waveform with frequencies in $\pm$2 GHz. The FPGA implements a high-throughput digital signal processing pipeline to down-convert each readout tone and monitor the phase. 

The RFSoC integrated ARM core runs Ubuntu-flavored Linux and serves as the command and control hub for each feedline readout. The processing system (PS) and programmable logic (PL) each have 4 GiB of available DDR4 SDRAM. The PL DDR4 serves as a data buffer for setup and calibration captures while the PS DDR4 is used for continuous photon capture and pixel monitoring. All FPGA capture, configuration, and control signals are managed by a Python package running on the PS. The PS is also able to program the IF board LO and attenuators to facilitate sweeps. 

The RFSoC4x2 platform has several quirks which impacted system design. The first was data converter synchronization. We found the phase mismatch between the DAC tile clocks disrupted the quadrature phase relation of our signals enough to cause -20 dB image tones. This did not meet our -30 dB IMD requirement (Sec. \ref{sec:req}) and forced us to implement Multi-Tile Synchronization (MTS)\cite{noauthor_multi-tile_nodate}. MTS suppresses the image tones to -40 dB but adds complexity and constraints to the clocking architecture.

The other quirk was the platform's default MPSoC clock configuration led the 1 Gigabit Ethernet to perform at roughly one-third capacity. We found the current 4x2 board support package (BSP) configures the full power domain main clock slower than the low power domain switch. We modified the BSP to correct this and recovered near line rate 1 GbE performance. More information on the Zynq UltraScale+ MPSoC is provided in reference\cite{noauthor_zynq_nodate}.

\subsection{\label{sec:fpga_design}FPGA Design}
The FPGA design followed a similar strategy to the Gen2 system but with several optimizations to manage double the number of channels and speed up data capture. The MKID readout waveform is computed in software and written to a large FPGA memory buffer where it can be replayed from the DACs in a loop. After the waveform passes through the feedline, it is sampled by the ADCs which feed into an intensive real-time processing pipeline. The digital signal processing steps are described in the following section.

\subsubsection{\label{sec:fpga_design}Digital Signal Processing}
\begin{figure}  
  \begin{center}
  \includegraphics[width=\columnwidth]{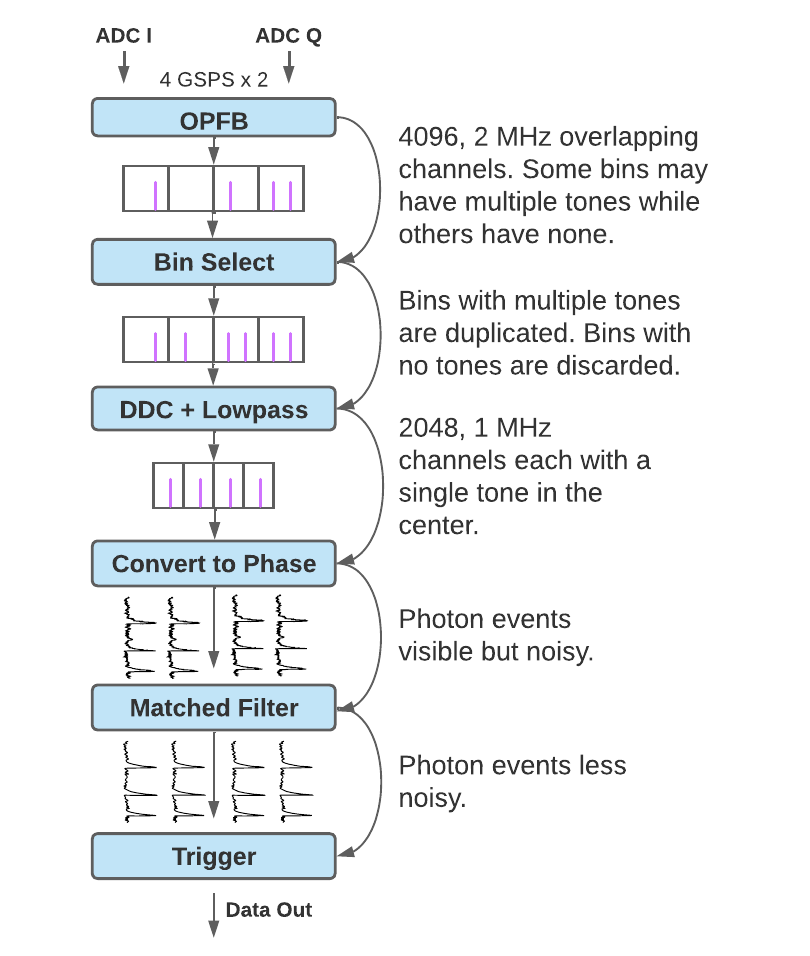}
  \end{center}
  \caption{Summary of key digital signal processing steps for frequency-division-multiplexed MKID readout. Signals are first channelized with a two-stage oversampled polyphase filter bank channelizer (OPFB) and direct digital-down-converter (DDC). The MKID tones are then filtered by custom matched filters before a trigger records photon time, energy, and location across all 2048 channels.}
  \label{fig:dsp}
\end{figure}

An overview of the key digital signal processing (DSP) steps is shown in Fig. \ref{fig:dsp}. After the 4 GHz waveform is injected into the programmable logic, it must be channelized into 2048 MKID channels. Due to fabrication uncertainties, the channelizer must be able to down-convert 2048 channels from arbitrary position in the microwave band. We use a two-stage oversampled polyphase filter bank (OPFB) channelizer followed by a direct digital-down-converter (DDC). The OPFB produces 4096, 2 MHz coarse channels which overlap 50\%. This allows every MKID readout tone to pass through the channelizer un-attenuated. The OPFB was highly-optimized as it is the most resource-intensive step and it is the subject of its own publication\cite{smith_high-throughput_2021}. 

After the OPFB, the Bin Select core selects the 2048 user-defined channels which contain MKID readout tones, copying bins with multiple tones as needed, and feeds them to the DDC. The DDC multiplies each channel by the complex conjugate of the readout tone to do the final down-conversion. The DDC core is also responsible for applying each channel's custom coordinate transform. The DDC complex multiplier includes an optional phase offset which can individually rotate each MKID loop. The core also implements a complex subtraction which serves to center each MKID loop. After each readout tone has been fully down-converted and the coordinate transform has been applied, each channel is low-pass filtered to remove other MKID readout signals that may fall in the same OPFB bin. The channels are decimated to produce 2048, 1 MHz fine channels each with one MKID readout tone in the center.

\begin{figure}  
  \begin{center}
  \includegraphics[width=\columnwidth]{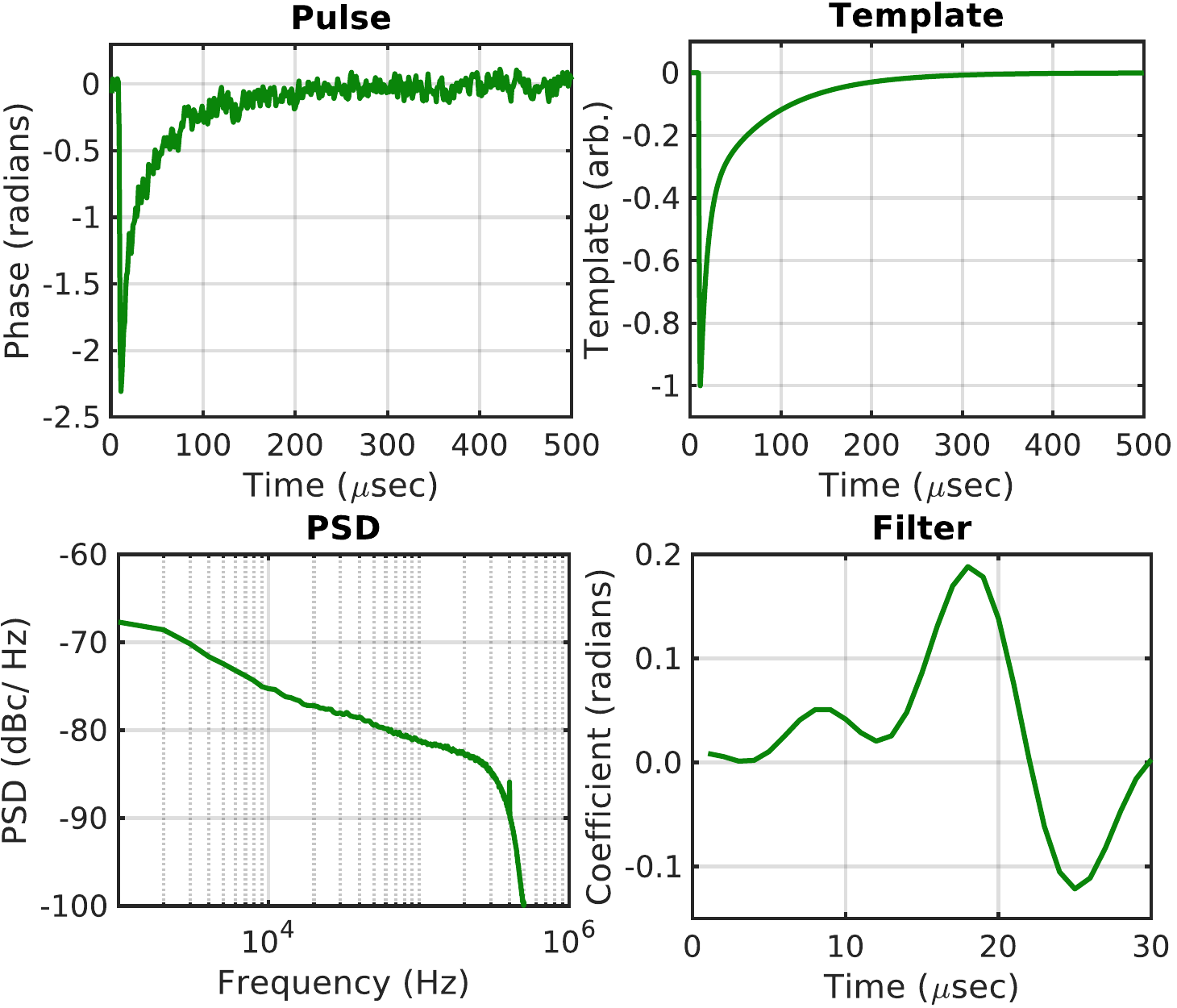}
  \end{center}
  \caption{Overview of MKID-specific matched filter generation. Photon pulses are recorded for each MKID pixel (upper left). Pulses are averaged in time to remove random noise and create a pulse template (upper right). Phase time streams from segments without photon pulses are averaged in the Fourier domain to estimate the phase noise in each channel (lower left). A Weiner filter is computed from the template and the noise and a final low pass filter is applied to produce each channel's matched filter (lower right).}
  \label{fig:optimal_filter_expl}
\end{figure}

The next step is to convert each channel to phase by evaluating $\mathrm{tan}^{-1}(Q/I)$. This operation results in 2048 phase time series each sampled every 1 microsecond. To further reduce phase noise, we apply a custom matched filter to every channel. The matched filter generation process in summarized in Fig. \ref{fig:optimal_filter_expl}. For every pixel on the feedline, we average unfiltered laser photons from the middle of the energy band to characterize the detector signal. We also estimate the channel phase noise by averaging the noise power spectral density when there are no photons arriving. With the measured signal and noise, we construct a matched filter for every channel.

\begin{figure}  
  \begin{center}
  \includegraphics[width=\columnwidth]{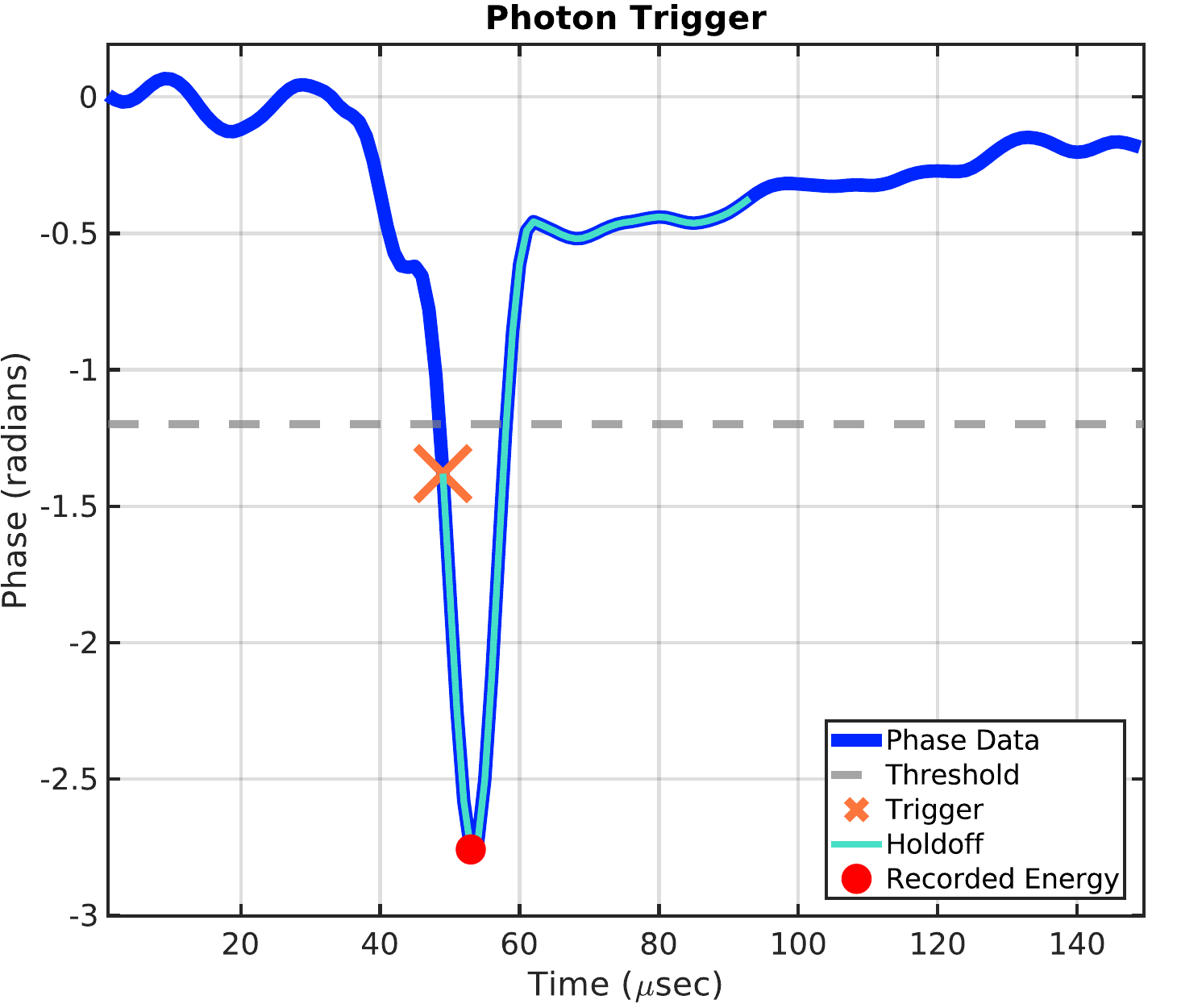}
  \end{center}
  \caption{Trigger operation on a filtered pulse. Once the phase crosses below the pixel's threshold, the channel triggers. During the trigger state, the minimum phase value is updated for some number of samples specified by the holdoff value. The minimum phase value in the holdoff window is recorded as the photon energy. After the holdoff expires, the channel is eligible to trigger again. Both the threshold and holdoff settings are calibrated per pixel.}
  \label{fig:trigger}
\end{figure}

After the matched filter enhances photon signals and suppresses channel-specific noise, we are finally ready for photon triggering. The trigger behavior is shown for a filtered pulse in Fig. \ref{fig:trigger}. Once the pulse crosses the threshold, the channel triggers. The minimum phase value is continually updated for a number of samples specified by the holdoff parameter. After the holdoff counter expires, the photon energy is recorded as the minimum value in the window and the channel is free to trigger again. The photon time, energy, and channel are recorded to disk.

\subsection{\label{sec:fpga_design}Implementation}
\begin{figure*}
\includegraphics[width=6.74in]{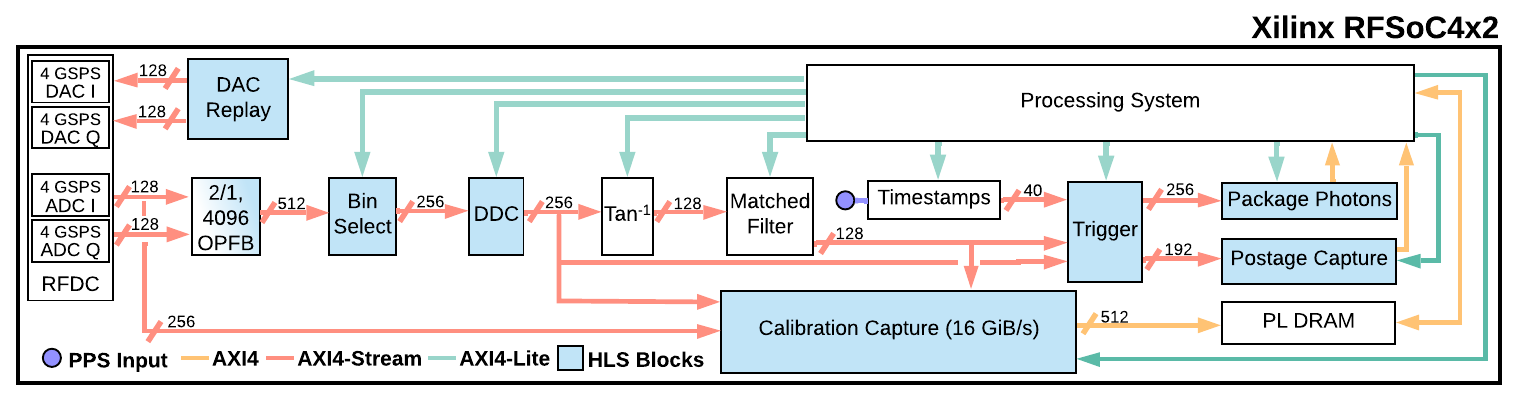}
\caption{\label{fig:block_design}Programmable logic system block design. Two 4.096 GSPS DACs produce the output waveform via a URAM lookup table. Two 4.096 GSPS ADCs feed data to a signal processing pipeline which monitors the phase of 2048 channels for photon events. A calibration capture subsystem routes streams to the PL DRAM, providing snapshots of data that can be used for setup and calibration tasks, such as identifying the MKID resonant frequencies, optimal readout powers, and phase biases. The processing system is used to configure blocks over AXI4-Lite and acts and the system command and control. Blocks implemented in Vivado/Vitis High-Level-Synthesis are shown in blue. Protocols and data-widths are labeled for select pathways.}
\end{figure*}

The FPGA design was implemented using Vivado 2021.1 in an IP integrator-based project flow. A system block design is shown in Fig. \ref{fig:block_design}. Resource utilization in summarized in Table \ref{tab:utilization} and device area utilization is highlighted by block in Fig. {\ref{fig:placement}}. The implementation is notable for realizing an intensive signal processing pipeline, requiring a large percentage of the FPGA resources at high clock rates, while using high-level synthesis tools. In this section, we will describe our approach to implementation, starting with our choice of tools.

\subsubsection{Tools}
All custom IP blocks, barring a few exceptions discussed below, were implemented using Vitis High-Level Synthesis (HLS) which synthesizes high-level C/C++ code to low-level hardware description language (HDL). HLS has the advantage of being more accessible to scientists without experience in hardware description language (HDL) and can be more flexible and easier to port to new hardware. However, it is notorious for using more resources and increasing timing strain. We developed several strategies for HLS usage to improve resource utilization and timing closure:

\begin{itemize}
    \item Small, single task, HLS blocks are preferable to complex blocks.
    \item Clean, small internal functions can significantly improve generated HDL.
    \item Partitioned temporary variables in unrolled loops is preferable to automatic inference within a loop.
    \item The \texttt{ap\_ctrl\_none} pragma directive can significantly improve control and logic optimization.
    \item  Manually picking bits instead of using the \texttt{DATAPACK} pragma directive can be vital.
    \item It is better to place memory resources such as BRAM and URAM manually using the IP generator in Vivado than to infer them in HLS.
    \item External AXIS to AXI conversion is necessary for continuous writes.
\end{itemize}

To help close timing in the full design, we used Vivado Intelligent Design Runs (IDR). IDR uses ML-based strategy predictions and incremental compile to help close timing in Vivado in tight designs\cite{noauthor_intelligent_nodate}. We used quality of result (QOR) suggestions and IDR-implemented runs to identify optimizations and achieve timing closure in the full design. Subsystem-specific strategies are discussed in context below.

\subsubsection{Waveform Replay \label{sec:dacreplay}}
The DAC Replay block streams a repeating waveform consisting of $2^{19}$ complex samples from a look-up-table (LUT) to the 4.096 GSPS DACs. The LUT size and DAC sample rate achieve 7.813 kHz frequency resolution for each superimposed tone in the readout comb. The table stores the 32-bit complex samples in a 2 MiB URAM buffer. An external URAM was used in Vivado as opposed HLS inference because HLS would not properly cascade the URAM blocks--a necessary optimization for reducing logic routing resources.

\subsubsection{Channelizer}
The OPFB, Bin Select, and DDC subsystems implement the channelizer. The OPFB is implemented using two HLS blocks that route and reorder data, sixteen parallel Xilinx FIR cores that implement the filter, and the Xilinx SSR FFT block exported from System Generator / Model Composer. Full implementation details are provided in \citet{smith_high-throughput_2021}. 

The Bin Select block caches eight copies of the preceding 4096-point FFT frame and routes them into eight parallel output lanes in any order (i.e., the pathological case of 2048 resonators in a single channel is supported). The channel order is run-time user programmable via AXI4-Lite. 

The DDC uses external BRAMs to store phase increments, offsets, and centers which are used with an internal cosine LUT to down-convert the channels and apply the complex rotation and translation coordinate transform. The DDC values are channel-specific and run-time user programmable over AXI4-Lite. A Xilinx FIR filter implements a low-pass and decimate operation to produce the fine channels.

\subsubsection{Convert to Phase and Filter}
The complex, demodulated signals are converted to phase through parallel Xilinx CORDIC\cite{noauthor_cordic_nodate} cores which implement $\mathrm{tan}^{-1}(Q/I)$. The matched filters are implemented using re-programmable Xilinx FIR compiler cores\cite{noauthor_fir_nodate}. The filter coefficients are run-time user programmable over AXI4-Lite.

\subsubsection{Photon Event Trigger}
The trigger is user-programmed at run-time with 2048 thresholds and holdoffs unique to each channel. The threshold is encoded as an 8-bit signed value allowing the phase threshold to be set with 0.02 radian precision. The holdoff value indicates the number of samples (microseconds) until a subsequent trigger is permitted. The holdoff takes integer values between 8 and 254. The maximum value of 254 corresponds to 254 microseconds which exceeds the typical filtered pulse recovery time (see Fig. \ref{fig:trigger}).  

\subsubsection{Photon Capture}
Photon capture is implemented using a pair of 800 KiB buffers. The processing system reads from one buffer while the other records photon data. The buffer is rotated either when full or when two photons arrive a user-defined interval ($\sim 0.0005 - 1$ s) after the first buffered photon.

\subsubsection{Postage Capture}
The postage capture system is capable of recording a 127-microsecond IQ time series window of up to 8000 total trigger events across any of 16 user-selected channels.

\subsubsection{Capture Subsystem}
\begin{figure}  
  \begin{center}
  \includegraphics[width=\columnwidth]{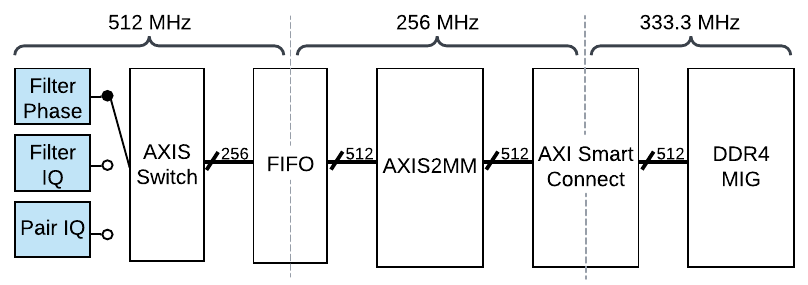}
  \end{center}
  \caption{Capture subsystem block design. Three HLS blocks (blue) package different streaming data formats from different places in the signal processing pathway. An AXIS switch selects the data stream to be captured. A dual-clocked FIFO feeds the selected data stream to an open source RTL core which produces AXI transactions for off-chip communication. The data is sent to the PL DDR4 via an AXI Smart connect and the Memory Interface Generator (MIG). The system is able to capture up to 4 GiB at 16 GiB/s.}
  \label{fig:capture}
\end{figure}

The capture subsystem manages data capture from various places in the signal processing pipeline to facilitate MKID feedline characterization and setup. An overview of the capture hierarchy is shown in Fig. \ref{fig:capture}. The system is backed by the PL DDR4 which provides 4 GiB of SDRAM storage. It is implemented using parallel HLS blocks that select user-requested groups of channels or, in the case of ADC capture, combine I and Q signals. A small, open-source RTL core\cite{noauthor_mazinlabwb2axip_nodate} is used to translate the AXI4-Stream interface to AXI4 transactions because HLS generated AXI4S-AXI4M interfaces do not support continuous write at the gigabyte level. AXI4 glue logic is used to cross clock domains and buffer the data before it is written to DDR4. The memory interface generator (MIG) is clocked at 333 MHz with a 512-bit bus, providing around >16~GiB/s offload bandwidth. 

\begin{figure}  
  \begin{center}
  \includegraphics[width=\columnwidth]{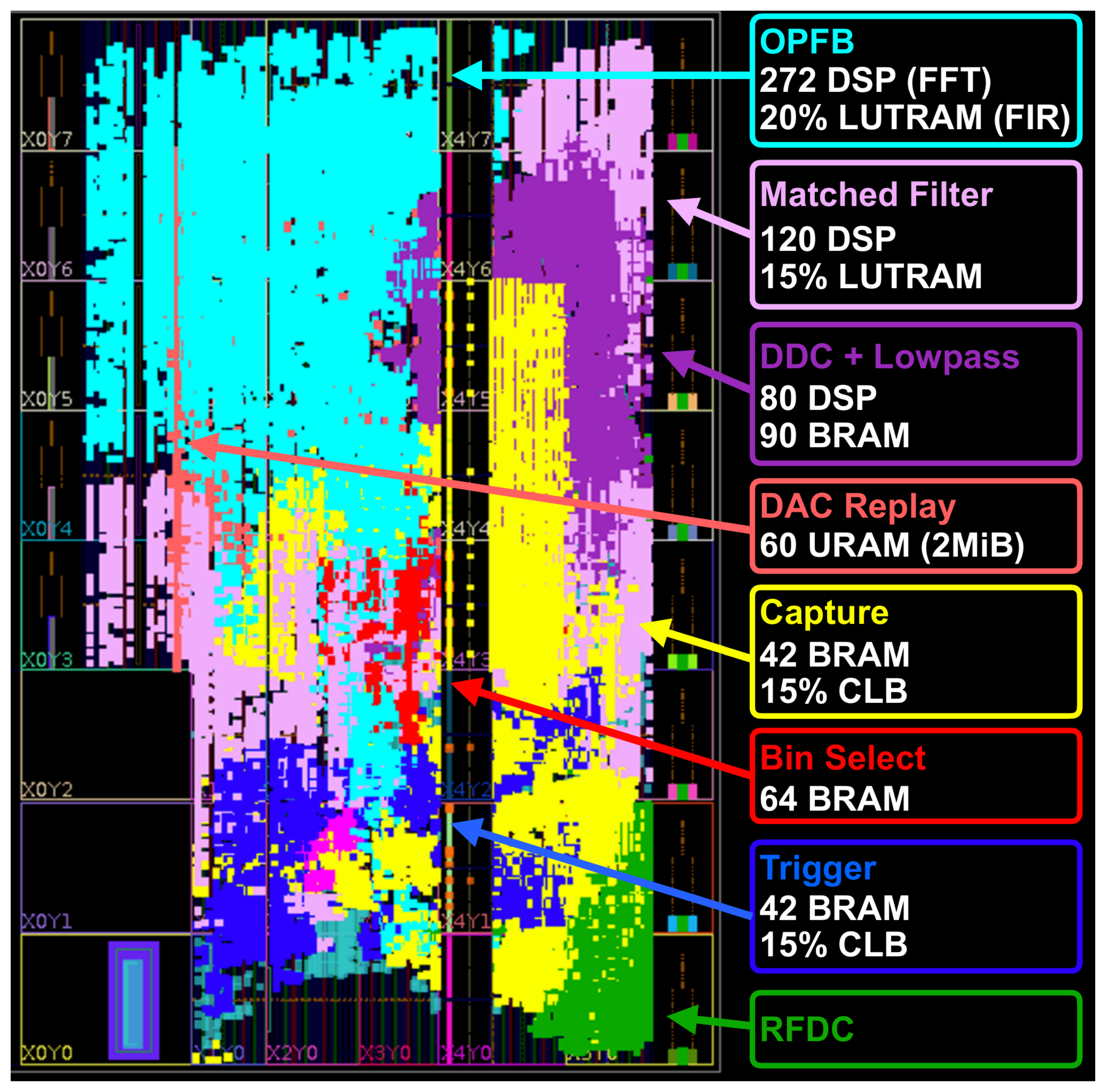}
  \end{center}
  \caption{Overview of the RFSoC ZU48DR device chip area utilization. Key signal processing subsystems are highlighted with different colors. Select FPGA resources are summarized for the highlighted systems on the right.}
  \label{fig:placement}
\end{figure}

Routing congestion caused by I/O pins to the PL DDR4 being proximate to the RFDC I/O pins caused trouble closing timing in the capture subsystem. A successful IDR run suggested the problem could be alleviated by constraining the MIG to the right side of the chip to create open lanes for the DAC output routing. The resulting square Pblock is visible in the placement view (see Fig. \ref{fig:placement}).

\subsubsection{Timestamps\label{sec:timestamps}}
Photon timestamps are provided by a time keeping core that supports several different modes. In basic operation, timestamps are generated using a 19-hour, 1 microsecond counter. All RFSoC boards are synchronized using the pulse-per-second (PPS) input linked to a GPS signal, providing a pulse each time the second rolls over. The current second as determined by UTC accessed from an NTP server by the processing system is synchronized with the PPS signal. The 36-bit, UTC-based timestamp is associated with each phase sample in the trigger block and recorded when a photon event is triggered. The resulting system provides 1~$\mu$s  timing resolution. Absolute timing error between boards was measured to be less than 150~ns\footnote{PPS signal was generated by a Stanford Research FS725 10 MHz Rubidium Frequency Standard.}.

\begin{table}
\caption{\label{tab:utilization}Resource utilization.}
\begin{ruledtabular}
\begin{tabular}{lcr}
Resource&Number&\% of ZU48DR\\
\hline
CLB & 36012 & 68\%\\
LUT (Logic) & 79809 & 19\%\\
LUT (Memory) & 98368 & 46\%\\
BRAM\footnote{Each BRAM is 36 Kbits.} & 262 & 24\%\\
URAM & 64 & 80\%\\
DSP & 475 & 11\%\\
\end{tabular}
\end{ruledtabular}
\end{table}

\subsection{\label{sec:software}Software}
The project software is based primarily on the Python Productivity for ZYNQ (PYNQ) framework. PYNQ is an open-source, AMD/Xilinx maintained project that facilitates interacting with and programming a ZYNQ architecture-based FPGA through Python\cite{noauthor_pynq_nodate}. PYNQ provides a Linux-based image, Jupyter notebook server, and extensible Python packages all running on the MPSoC. The \texttt{mkidgen3} package includes Python drivers for the FPGA design IP, MKID-specific setup and calibration functions, and diagnostic plots. In the following subsections we detail our adoption and adaptation of the PYNQ framework to support MKID readout.

\subsubsection{Deviations from the PYNQ Image}
The \texttt{mkidgen3} system image has two main modifications from the Ubuntu 22.04-based PYNQ 3.0.1 stock image. First, we reserved a segment in the device tree for the PL DDR4. Editing the device tree allowed more robust, full integration of the PL DDR4 with the PYNQ memory model and recommended \texttt{pynq.allocate} method. The second adaptation was to patch the \texttt{xrfdc} Bare Metal C RFDC driver to include MTS-related functions. These deviations are documented in the main project repository.

\subsubsection{\label{sec:rfsoc4x2} Python Drivers}
PYNQ provides a simple methodology to create Python drivers for IP. The framework parses the Vivado-generated hardware hand-off (\texttt{.hwh}) file, effectively auto-discovering IP, and provides the user with a \texttt{pynq.Overlay} object that can be used to command and control design IP through its associated register map. We use this approach to write Python drivers for all configurable IP in the design. The PYNQ drivers can be used directly to set up and read out MKIDs using the on-board Jupyter notebook server or through remote Python execution, i.e.,  over \texttt{ssh}.

\subsubsection{\label{sec:rfsoc4x2} Server Architecture}
The software ecosystem also includes high-level client and server software that facilitate array-level set up, capture, and observations and is intended for use in a multi-board MKID instrument environment. Commands are based on the \texttt{zmq} request-reply pattern with each RFSoC running a feedline server that schedules capture requests from the client and guarantees all captures occur with the required FPGA state. Resulting capture data and status updates are broadcast via the \texttt{zmq} pub-sub pattern, allowing any network device to monitor photon or engineering data. At the time of writing, the server is still undergoing testing and will be fully detailed in a future publication.

\section{\label{sec:perf}System Performance Characterization}
\begin{figure}  
  \begin{center}
  \includegraphics[width=\columnwidth]{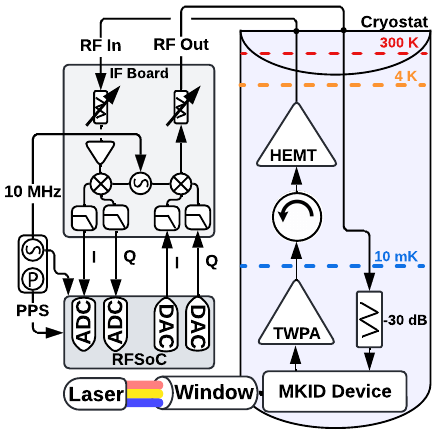}
  \end{center}
  \caption{Experimental setup for system performance characterization. RFDC loopback measurements were taken with the RFSoC DAC \textit{I} and \textit{Q} connected to the RFSoC ADC \textit{I} and \textit{Q} ports via external SMA cables (not shown). IF loopback measurements were taken with the IF board \textit{RF Out} connected to the IF board \textit{RF In} with a 30 dB fixed attenuator in-between the ports (not shown). MKID measurements were taken with the setup shown and include the cryogenic wiring and amplifier chain. During MKID testing, lasers are shone through a window in the cryostat to illuminate the device with known-energy photons.}
  \label{fig:white_fridge}
\end{figure}

The Gen3 MKID readout system has been successfully demonstrated in lab and meets all performance requirements outlined in Sec. \ref{sec:req}. The system was characterized in three configurations:
\begin{enumerate}
    \item \textbf{RFDC Loopback:} The RFSoC4x2 I and Q ADCs are directly connected to the I and Q DACs, respectively, with short, matched cables.
    \item \textbf{IF Loopback:} The IF board \textit{RF Out} is connected to the IF board \textit{RF In} with a 30 dB fixed attenuator in-between.
    \item \textbf{MKID Measurement:} The full system is connected to a cryogenic UVOIR MKID device (see Fig. \ref{fig:white_fridge}).
\end{enumerate}
In all configurations, a Stanford Research FS725 10 MHz rubidium frequency standard was used to synchronize the RFSoC4x2 and IF board clocks and provide the PPS signal. 

\subsection{Multi-MKID Readout Approximation\label{sec:randcomb}}
A pseudo-random comb was used to simulate the simultaneous readout of 2048 MKIDs. The comb consists of a superposition of readout tones where each tone is randomly placed in every other 2 MHz coarse channel in the $\pm$ 2 GHz readout band for a total of 2048 approximately-evenly-spaced tones. The tones are designed to be no closer than 300 kHz to their 2 MHz channel edge which mimics the OPFB overlap region and simplifies channel assignment in the FPGA. The resulting waveform is similar to what we find with modern UVOIR MKID devices and provides realistic conditions for evaluating multi-MKID readout performance.

\subsection{\label{sec:combperf} Output Comb Performance}
\begin{figure}  
  \begin{center}
  \includegraphics[width=\columnwidth]{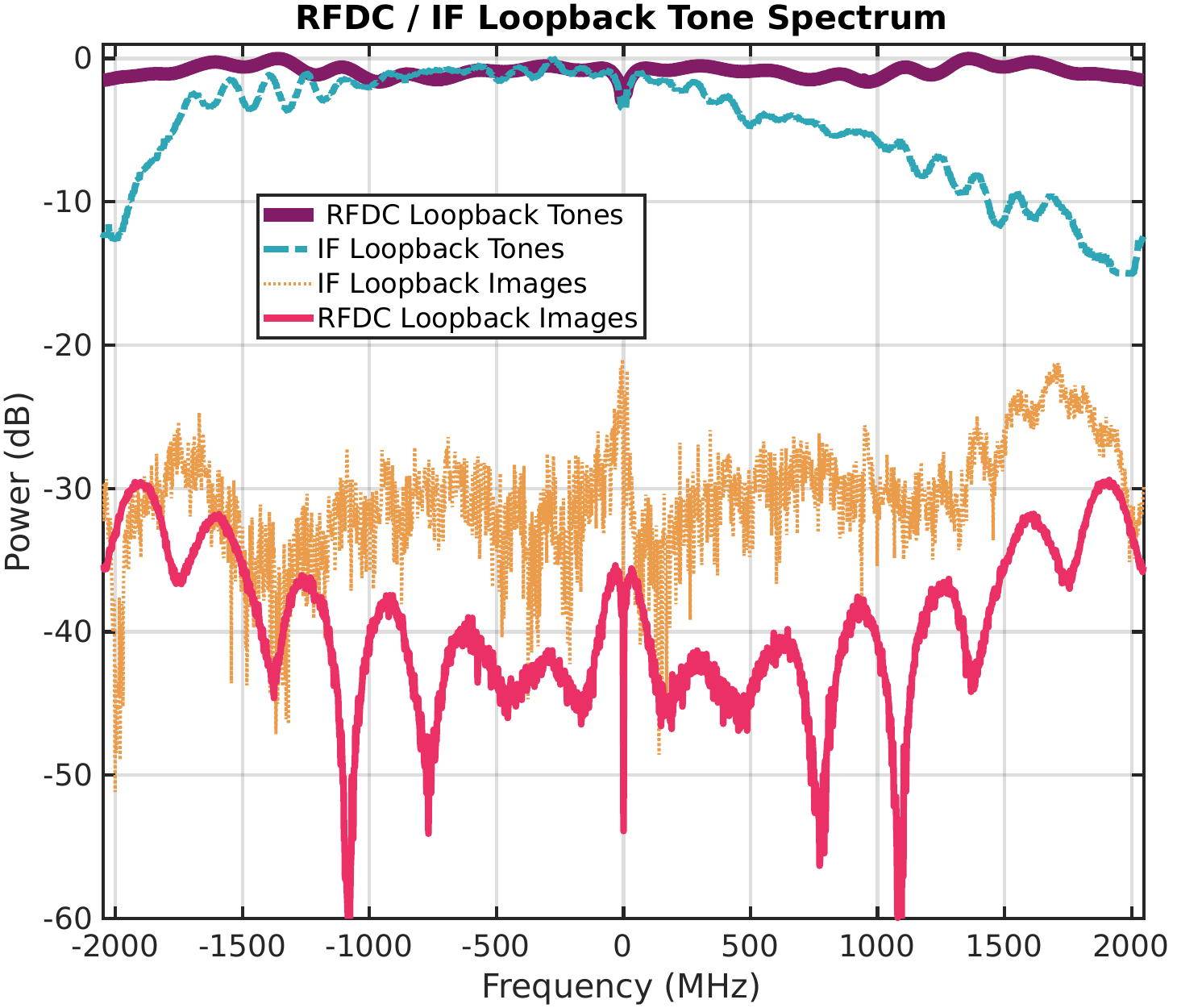}
  \end{center}
  \caption{Power level of 2048 driven readout tones taken with the system in RFDC loopback (solid purple) and IF loopback (dashed blue). Power level of non-ideal images tones created by IQ imbalance are shown for IF loopback (dotted gold) and RFDC loopback (solid red). As expected, the IF loopback traces show more tone power loss and higher image tones due to IQ imbalance and impedance mismatches. The effects of the IF board anti-aliasing filters are also visible as attenuation at the edges of the band in the IF loopback tone trace. }
  \label{fig:adc_snap}
\end{figure}

A 2048-tone, uniform-amplitude, pseudo-random-frequency (see Sec.\ref{sec:randcomb}) comb was generated and played out the DACs to simulate an MKID readout waveform. The output tone and image powers for all frequencies in IF loopback and in RFDC loopback are shown in Fig. \ref{fig:adc_snap}. In RFDC loopback, the comb is fairly uniform, showing only a few dB variation across the $\pm$ 2 GHz readout band. The image tones are at most -30 dB at the band edges and meet our performance requirement (Sec. \ref{sec:ctimd}). In the IF loopback, the output tone power is more attenuated and non-uniform across the band. The attenuation at the edges of the 4 GHz band are caused by imperfect roll-off in the IF board analog anti-aliasing low-pass filters (see Sec. \ref{sec:if}). Ripples in the tone power are caused by impedance mismatches which launch reflections and creating standing waves in the IF board. The real-valued frequencies (0, 2 GHz) are more attenuated than the imaginary-valued frequencies (-2 GHz, 0), likely due to excess attenuation in the IF board I signal path. We hypothesize the manual rework of the anti-aliasing filters in-house resulted in poor microwave hygiene and wide variation between signal paths in the IF board, leading to reduced performance.

\subsection{\label{sec:loopback_phase_noise} Phase Noise}
\begin{figure}  
  \begin{center}
  \includegraphics[width=\columnwidth]{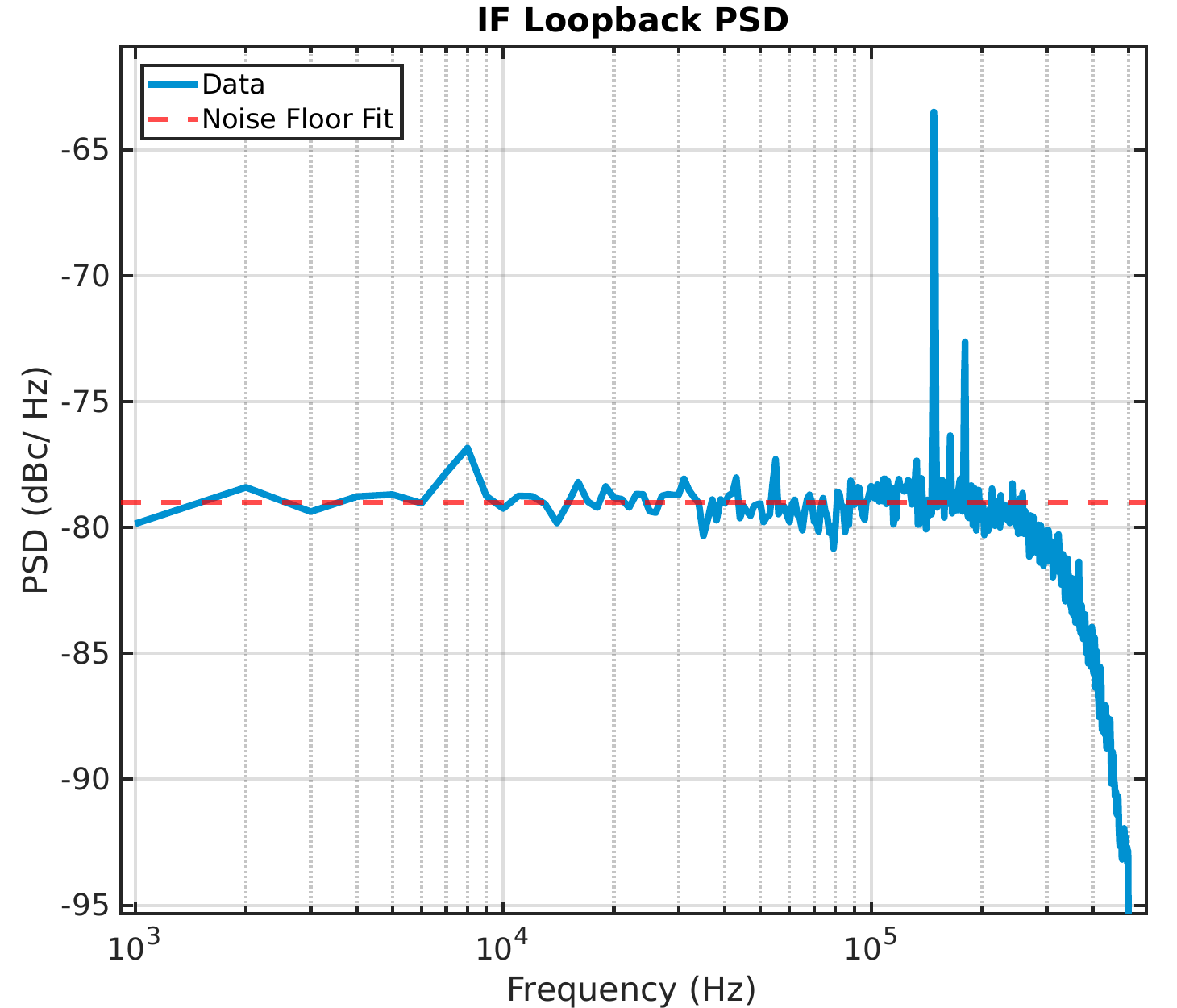}
  \end{center}
  \caption{Phase noise power spectral density in a single channel. No matched filter is applied. The data was collected with the system in IF loopback with 2048 pseudo-random tones. The white noise floor is fit with the red dashed line.}
  \label{fig:single_channel_psd}
\end{figure}

The phase power spectral density from a down-converted readout tone is shown in Fig. \ref{fig:single_channel_psd}. The phase noise floor is fit with the red dashed line and is approximately -80 dBc/Hz. There is some line noise present around 150 kHz, possibly from electromagnetic interference in the lab or a nearby image tone. Line noise can also be generated from rounding errors in the DSP pipeline which uses truncation as the default rounding mode to conserve FPGA resources. Truncation can introduce a DC bias, for example in an OPFB channel, which may be up-converted to an image tone during the DDC multiply operation.  

\begin{figure}  
  \begin{center}
  \includegraphics[width=\columnwidth]{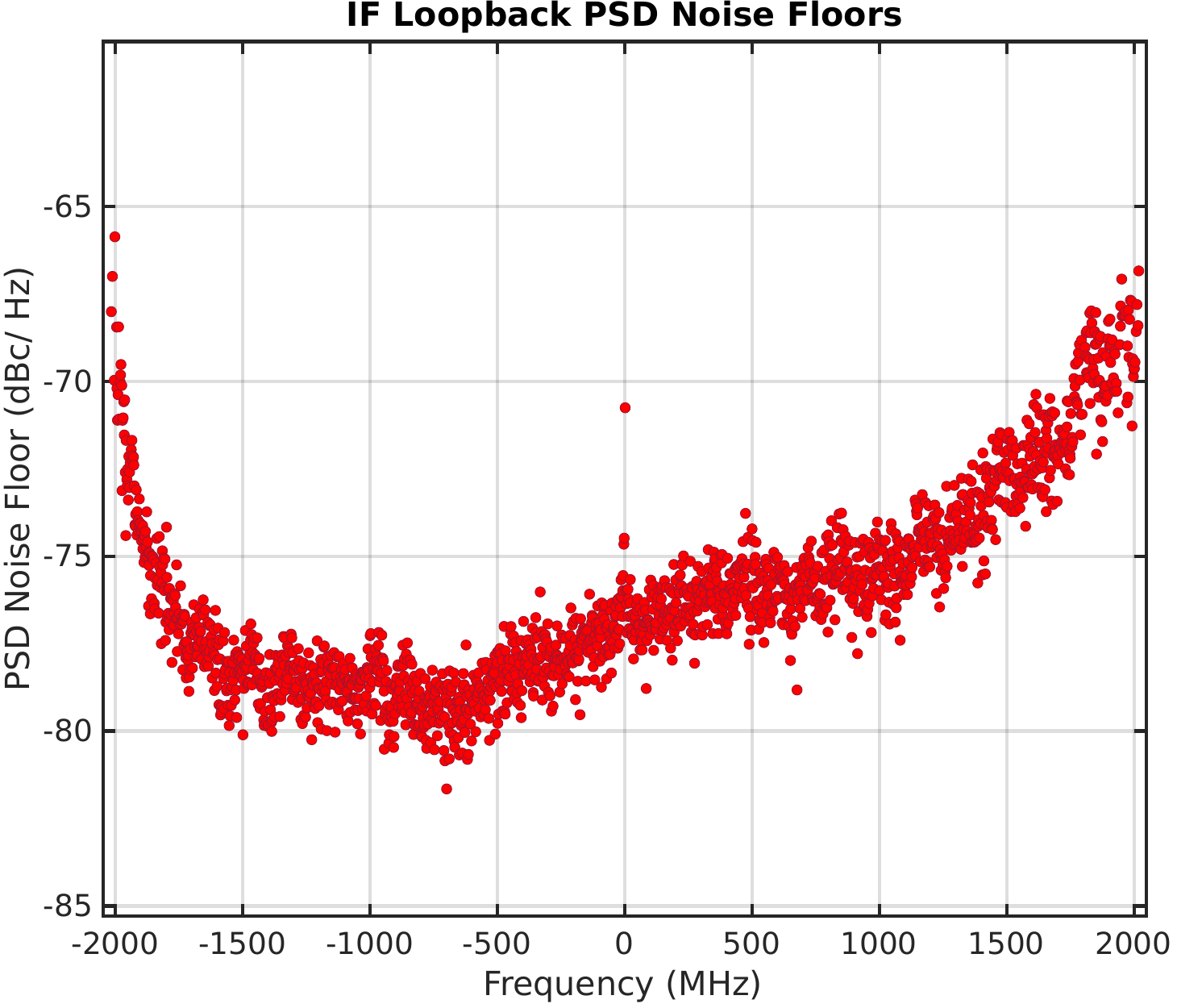}
  \end{center}
  \caption{Phase noise power spectral density noise floor fit in all channels. The data was collected with the system in IF loopback with 2048 pseudo-random tones running. The white noise floor was fit and recorded for each channel as shown in Fig \ref{fig:single_channel_psd}. The phase noise drifts $\sim$10 dB across the band, likely due to IQ imbalances in the system.}
  \label{fig:psd_floors}
\end{figure}

Fig. \ref{fig:psd_floors} shows the PSD-floor fit for all 2048 channels in the pseudo-random-tone waveform. The PSD floors follow a pattern inverse to the IF loopback tone power measured in Fig. \ref{fig:adc_snap}, with the real-valued frequencies showing more noise. This pattern is indicative of the same issue with the IF board microwave hygiene and poor transmission performance in the \textit{I} signal path. 

\subsection{\label{sec:mkidreadout} MKID Readout}
The Gen3 readout system was also characterized in lab using a real MKID and a series of lasers. These studies provide a concrete means to study the system's energy-resolving capability and help disentangle effects of readout-specific noise on scientific utility.

\subsubsection{Experimental Setup}
The MKID device was a PtSi-on-sapphire, array-style chip\footnote{Full fabrication details are provided in \cite{szypryt_large-format_2017}.} with the same design and manufacturing process used for the device in MEC and in \citet{zobrist_wide-band_2019}. The MKID was cooled in a dilution refrigerator (20 mK) with a first-stage, quantum-noise-limited parametric amplifier\cite{faramarzi_4-8_2024}. Lasers at 405.9 nm, 663.1 nm, and 808.0 nm illuminated the MKID through windows in the the dilution refrigerator. The experimental setup is shown in Fig. \ref{fig:white_fridge}. 

A single 4.5 GHz MKID was characterized, biased, and prepared for photon readout using the \texttt{mkidgen3} Python package on the RFSoC-hosted Jupyter Notebook. The resonator was set up manually using a series of sweeps and was ultimately biased at the correct frequency using the IF board programmable LO, and the correct power using the IF board \textit{RF Out} programmable attenuator. This technique is not suitable for setting up multiple MKIDs on one feedline because the LO and programmable attenuator provide global frequency and power offsets; however, it provides a decent approximation for the full-setup algorithms and server utility that are currently are under development. To characterize the extra readout noise generated during multi-MKID readout, excess readout tones were pseudo-randomly generated (see Sec. \ref{sec:randcomb}) and propagated through the MKID device and full readout chain.

For maximum flexibility in the data analysis, the raw phase time series was captured with a unity matched filter (no effect) and the matched filter and trigger algorithms were applied in software. The matched filter and trigger performance were verified on the FPGA for a subset of experimental settings.

\begin{figure}  
  \begin{center}
  \includegraphics[width=\columnwidth]{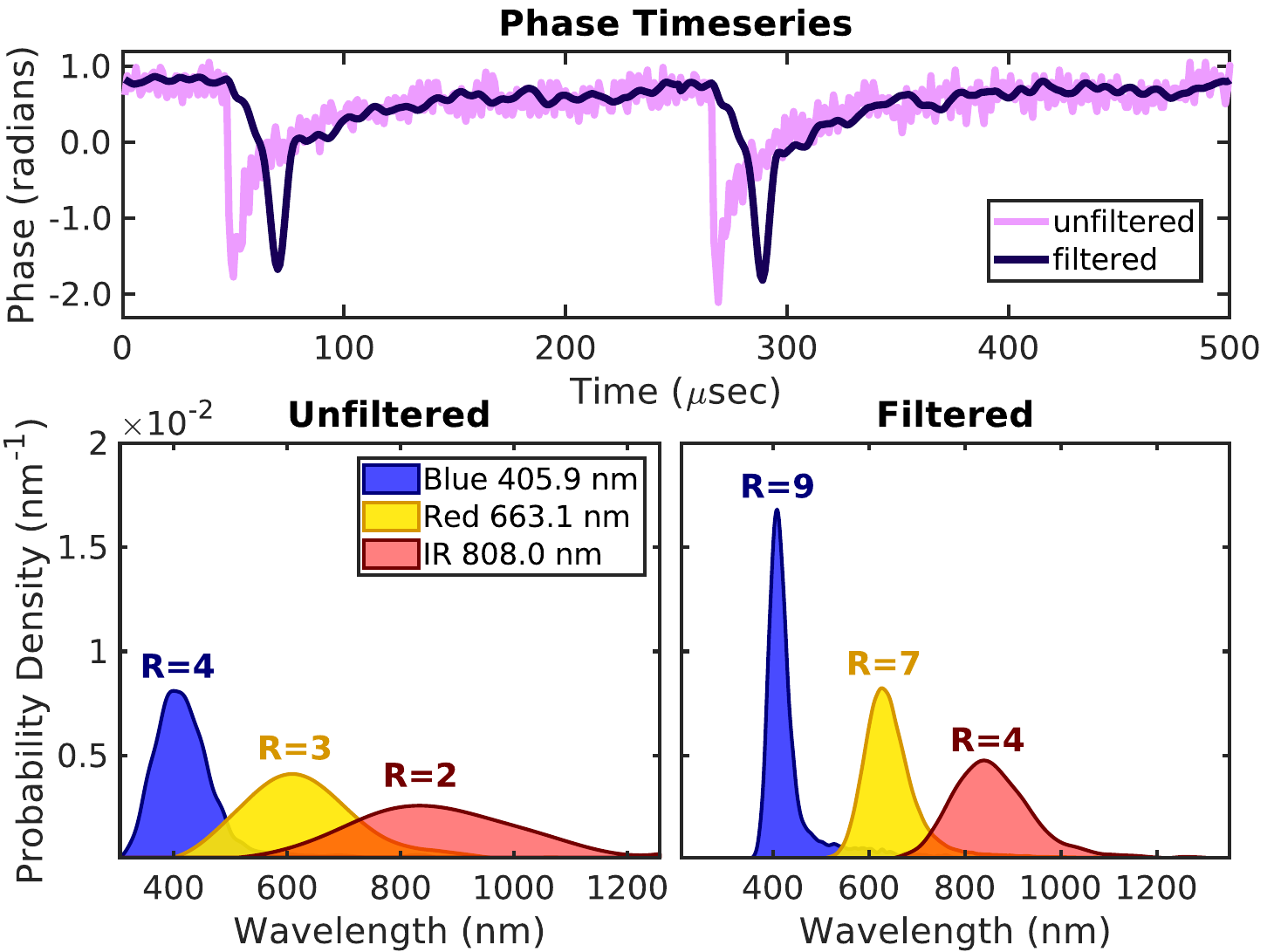}
  \end{center}
  \caption{Summary of matched filter performance in the case of 2048 pseudo-random tones. Top: Phase time series taken with a red 663.1 nm laser illuminating the MKID without (light pink) and with (dark pink) the 30-tap matched filter applied in software. Bottom: Resolving power achieved without (left) and with (right) the matched filter applied. The resolving power, $R$, is annotated above each energy histogram. Different wavelength photons are shown in different colors. The matched filter reduces noise in the phase time series and greatly improves resolving power.}
  \label{fig:ofilt_result}
\end{figure}

\subsection{Matched Filter Performance}
The matched filters are chiefly responsible for boosting the signal-to-noise ratio in the phase time series and improving resolving power. The test MKID's 30-tap filter was computed as described in Fig. \ref{fig:optimal_filter_expl} using pulse data from the middle (red, 633.1 nm)  laser wavelength. The detector resolving power was characterized with and without the matched filter. In each case, the readout system used a 2048-tone, pseudo-random comb to simulate the noise environment of a full-feedline readout. Results are summarized in Fig. \ref{fig:ofilt_result} and show a dramatic improvement in resolving power with the matched filter applied. 

\begin{figure*}
\includegraphics[width=6.74in]{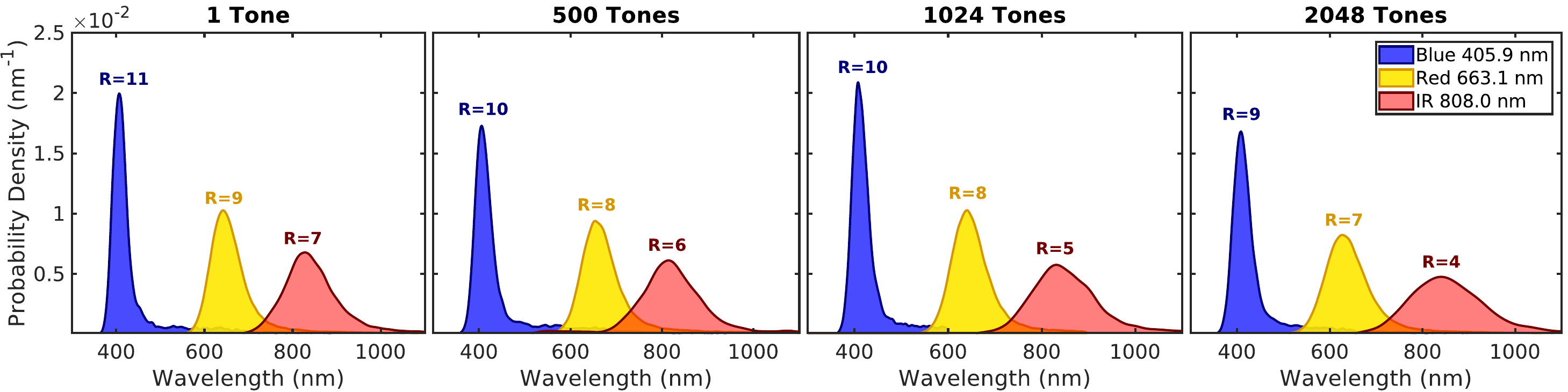}
\caption{\label{fig:multi_tone}Single MKID response to different wavelength photons as more pseudo-random tones are added to the readout comb. The resolving power, $R$, is annotated above each energy histogram. Different wavelength photons are shown in different colors. The resolving power degrades slightly as more tones are added.}
\end{figure*}

\begin{figure}  
  \begin{center}
  \includegraphics[width=\columnwidth]{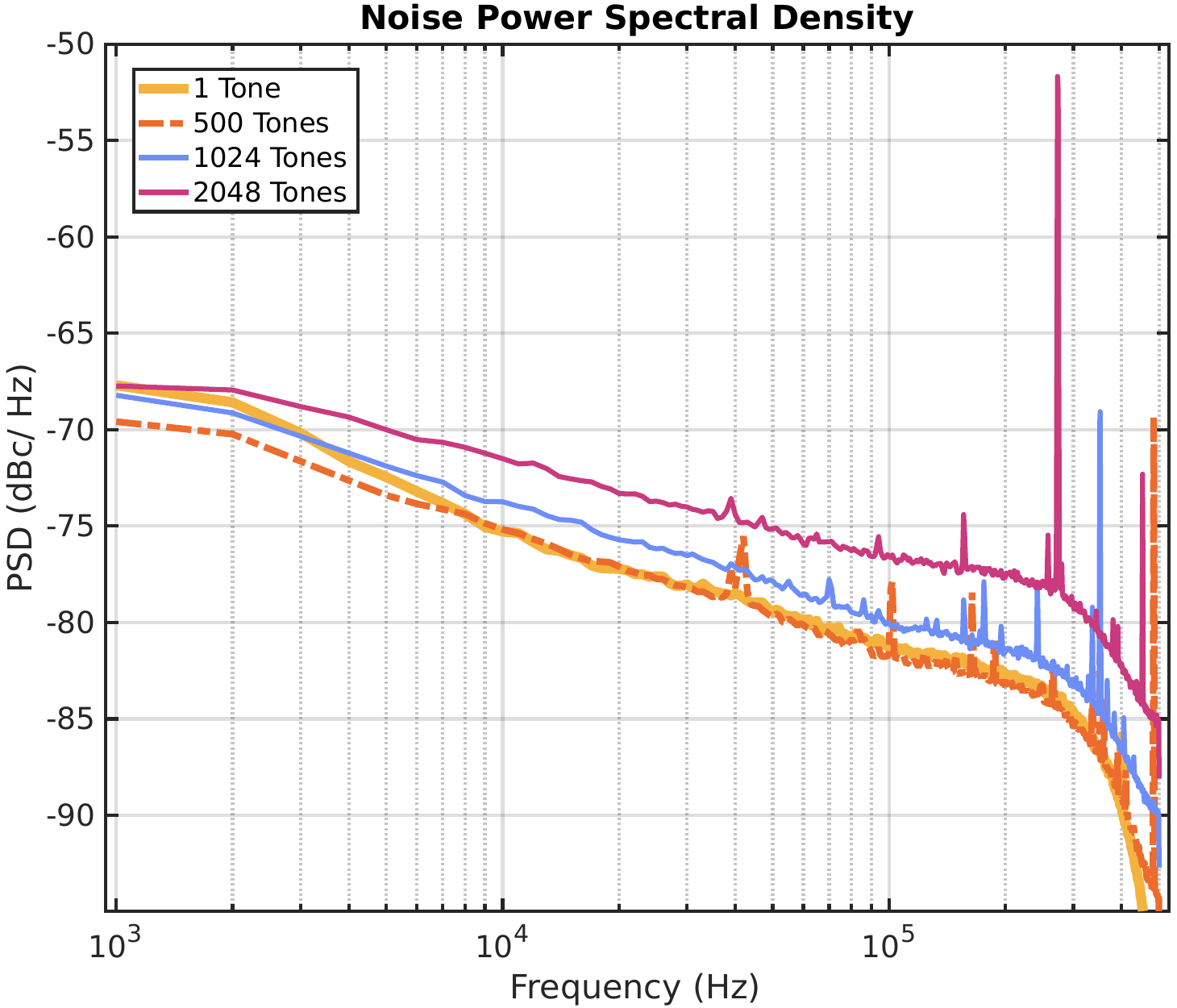}
  \end{center}
  \caption{Phase noise power spectral densities taken with the system connected to the fridge as shown in Fig. \ref{fig:white_fridge}. The data was collected with the readout tone biased to the optimal MKID resonance frequency and power and includes MKID device two-level system noise as well as noise from the cryogenic amplifier chain. As more tones are driven, the phase noise floor and spurious line noise both increase. When 2048 tones are driven, a large tone appears around 280 kHz at the -52 dB level; it may be an image tone or intermodulation product from one or more of the other driven tones.}
  \label{fig:noise_psd}
\end{figure}

\begin{figure}  
  \begin{center}
  \includegraphics[width=\columnwidth]{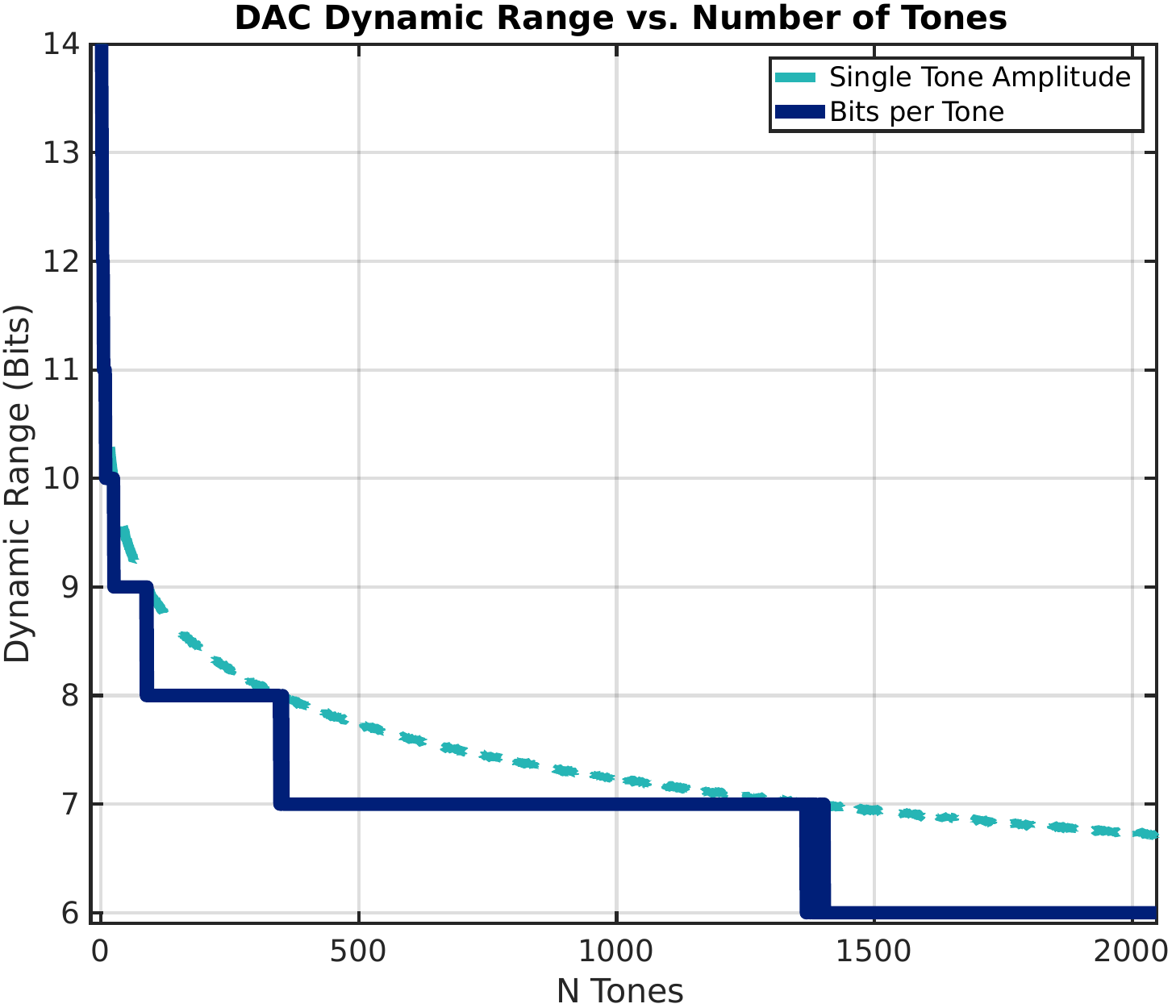}
  \end{center}
  \caption{Bits of DAC dynamic range available per readout tone as a function of the number of tones assuming all tones have equal amplitudes and random phases. The light blue dashed line represents $\mathrm{log_{2}}$ of the average individual tone quantized amplitude. The actual number of bits is the $\mathrm{floor}$ of this result and is shown in solid navy. There's a large fall-off in DAC dynamic range between a single tone and 500 tones. At our maximum number of channels, there are six bits available per tone.}
  \label{fig:dac_dr}
\end{figure}

\subsection{Resolving Power Results}
The detector resolving power, $R$, was measured with different numbers of pseudo-random tones in the readout waveform to characterize the best-case and worst-case scenario performance. Results are summarized in Fig. \ref{fig:multi_tone}. In the case of only one readout tone in the waveform, we find excellent resolving power that is consistent with \citet{zobrist_wide-band_2019} and meets our best-case scenario performance requirement (Sec. \ref{sec:rp}). In the worst-case scenario where we are driving all 2048 tones, we see some degradation in the resolving power but the results are within the acceptable limits set by the Gen2 MEC deployment (Sec. \ref{sec:rp}).

\subsection{Resolving Power Degradation with Tone Number\label{sec:multi_tone_deg}}
While the resolving power requirements are met, future MKID devices are expected to have better intrinsic resolving power. Understanding the nature of the resolving power degradation is useful for planning system upgrades and drafting requirements for future systems. The MKID phase noise power spectral density for each number of tones is shown in Fig. \ref{fig:noise_psd}. The data was collected with the readout tone biased to the optimal MKID resonance frequency and power and includes the MKID device two-level system noise as well as noise from the cryogenic amplifier chain. Fig. \ref{fig:noise_psd} shows that as the number of tones goes up, both the noise floor and line noise increase.

We hypothesize there are three main ways in which increasing the number of tones in the waveform can raise system noise and decrease resolving power. 
\begin{enumerate}
    \item Reduced DAC Dynamic Range

    For best noise performance, the DAC maximum voltage output is typically mapped to the maximum value of the readout waveform. The IF board \textit{RF Out} programmable attenuator can be used to reduce the power level to the optimal MKID drive power (around $-100$ dBm) as needed. As more tones are added, the maximum value of the superimposed readout waveform grows and there is less available DAC dynamic range per readout tone. The reduction in number of bits available per tone is shown in Fig. \ref{fig:dac_dr}. This effect increases the quantization noise in each channel.

    \item Reduced ADC Dynamic Range
    
    Similar to the DAC, there is also a reduction in ADC dynamic range per tone as more tones are added. Typically, the IF board \textit{RF In} programmable attenuator is adjusted so the maximum waveform value will be close to the maximum ADC input voltage. The more tones superimposed in the waveform, the fewer bits available per tone. A similar relation to Fig. \ref{fig:dac_dr} applies except the RFSoC4x2 ADC maximum dynamic range is 12-bits, not 14.

    \item Intermodulation Effects

    Finally, there are spurs, intermodulation products, image tones, and reflections all caused by the proliferation of readout tones in the analog domain. These signals may also impact dynamic range in the digital domain, for example, more noise power will cause more signal resolution to be lost during truncation rounding.
\end{enumerate}

\begin{figure}  
  \begin{center}
  \includegraphics[width=\columnwidth]{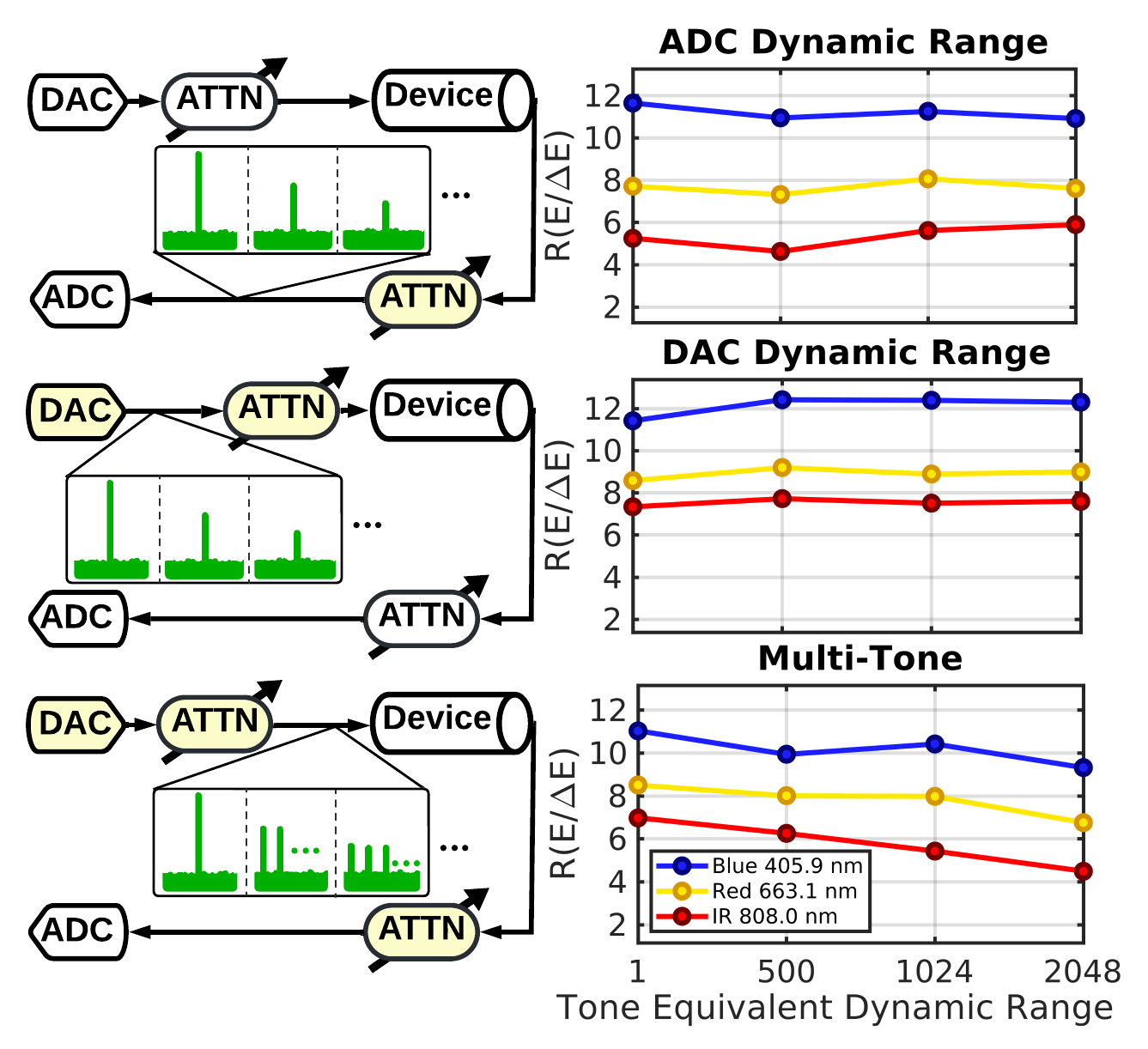}
  \end{center}
  \caption{\label{fig:rlines} Three experimental setups (left) and results (right) intended to disentangle the ADC and DAC dynamic range from intermodulation product effects on detector resolving power. In each experiment, changing components are highlighted in yellow. (Top) A single readout tone is output by the DAC at a dynamic range equivalent to that available when driving 2048 tones (see Fig. \ref{fig:dac_dr}). The ADC programmable attenuation is increased to lower the ADC dynamic range to that available per tone with 1, 500, 1024, and 2048 tones. The ADC dynamic range does not appear to have a large impact on resolving power. (Middle) The ADC input attenuator is fixed at a level compatible with 2048 tones while the DAC outputs a single tone with various dynamic range utilization as indicated by the number of tones. The DAC programmable attenuator is changed with each step to compensate for the reduced waveform amplitude and keep the readout tone power at the correct bias level for the MKID device, isolating the DAC dynamic range. The DAC dynamic range does not show a large impact on R at the levels tested. (Bottom) The impact of driving more tones on resolving power. The DAC and ADC programmable attenuators are adjusted with each number of tones to use the full dynamic range of both data converters. Histograms for this study are also shown in Fig. \ref{fig:multi_tone}.  We observe some degradation in R with increasing tone count, suggesting that intermodulation products and image tones arising from the multitude of driven tones are more damaging to resolving power than the increase in quantization noise from data converter dynamic range constraints.}
\end{figure}

We devised two additional studies to isolate the ADC and DAC dynamic range effects and determine the leading contributor to the resolving power degradation we observe with increasing tone number. In the first study, a single tone was played from the DAC but the waveform was artificially made to use different DAC dynamic ranges in accordance with the values expected for different numbers of tones (see Fig. \ref{fig:dac_dr}). For each DAC dynamic range setting, the \textit{RF Out} IF board programmable attenuator was adjusted to ensure the MKID device was still being driven at the correct level. The ADC input attenuator was fixed conservatively at a level consistent with what would be compatible with 2048 tones. In this way, we were able to probe the effects of the loss in DAC dynamic range absent intermodulation products and changes in ADC dynamic range utilization. To study the effects of ADC dynamic range, we again played a single tone from the DAC but this time at a fixed dynamic range, again conservatively in-line with what would be available with 2048 tones. The IF board \textit{RF In} programmable attenuator was adjusted to purposefully lower the ADC dynamic range used to what would be available per tone for different numbers of tones. Results for these studies is compared with the data collected for the multi-tone experiment in Fig. \ref{fig:rlines}. The experiments show that the degradation in DAC and ADC dynamic range, and consequently the increase in quantization noise, does not significantly impact device resolving power. In all cases, the measured resolving power is similar to what was measured in the best-case scenario with a single tone and maximal ADC/DAC dynamic range utilization. We conclude that spurious tones, intermodulation products, image tones, and reflections caused by the multitude of readout tones is the primary factor driving resolving power degradation with increasing number of tones.

\section{Discussion\label{sec:discussion}}
The Gen3 system characterized in this work is capable of reading out 2048 UVOIR MKIDs across 4 GHz of bandwidth using an RFSoC and IF board. While the system meets all performance requirements outlined in Sec. \ref{sec:req} and provides a major improvement in UVOIR MKID readout scalability, there are several areas that could be improved and questions that warrant further investigation. Performance-based trade-offs and recommendations are discussed in context below.

\subsection{IF Board IQ Imbalance and Images}
We use quadrature sampling in the RFSoC along with IQ mixers in the IF board to access the 4-8 GHz UVOIR MKID band. Quadrature sampling has the benefit of reducing the required data converter speed by half in order to access a given band. At present, this technique is necessary to reach our desired readout band using an RFSoC platform. Unfortunately, quadrature sampling comes at the cost of being very sensitive to gain/phase imbalance between the two quadratures which results in undesired image tones. These imbalances can be corrected at a single frequency but the calibration complexity grows significantly with bandwidth and in practice it is not feasible to calibrate more than about 400 MHz of instantaneous bandwidth\cite{noauthor_image_nodate}.

In the RFSoC, using Multi-Tile Synchronization\cite{noauthor_multi-tile_nodate}, we achieve enough consistency between the I and Q paths that the image tones are within the -30 dB requirement (Sec. \ref{sec:ctimd}); however, this is not the case when the IF board is included (see Fig. \ref{fig:adc_snap}), with images rising to within -25 dB and even -15 dB in some cases with respect to the driven tones. As discussed in Sec.\ref{sec:combperf} and Sec. \ref{sec:loopback_phase_noise}, the IF board anti-aliasing filters were manually reworked in lab, resulting in poor signal integrity and imbalance between the I and Q paths (see Fig. \ref{fig:adc_snap} and Fig. \ref{fig:psd_floors}). While the current IF board prototype provides a means to measure MKIDs in the lab, it does not meet the performance criteria across the full readout band and is likely not suitable for instrument deployment. With a modest redesign and professional PCB manufacturing, we expect the IF board loopback measurement can achieve $\le$ 30 dB image tones and uniform -80 dBc / Hz phase noise floor across the band. 

\subsection{Direct RF Sampling}
Even better than redesigning the IF board would be to do away with it entirely and use a direct RF approach. This eliminates the need for quadrature sampling with separate I and Q signal paths, potentially greatly improving RF signal integrity, readout phase line noise, and ultimately detector resolving power. Current RFSoC devices do not have the data converter speed required to sample 4-8 GHz directly in either the first or second Nyquist zone. It may be possible to use the 5.9 GSPS DFE device\cite{noauthor_zynq_nodate} at the maximum rate in the third Nyquist zone but more testing is needed to verify performance. Another approach is to use the second Nyquist zone (2.95-5.9 GHz) and modify the MKID readout band, though this would require significant device re-design and may not be compatible with commercial, low-noise cryogenic amplifiers required to boost the device signal. A faster RFSoC may be available soon in the Versal AI RF part which is advertised as having 36 GHz of analog bandwidth (4 GHz IBW per channel) and would likely be able to directly access the full readout band using the first or second Nyquist zone\cite{noauthor_aerospace_nodate}. 

With direct-RF-capable platforms on the horizon, the Gen3 system design prioritized platform-upgradability. The modular approach disentangles I/O from core DSP blocks and is backed by a C \texttt{Makefile} and Vivado \texttt{tcl} scripts that that should greatly simplify re-targeting the design to a new RFSoC. The HLS DSP blocks expedite updates by exposing bit-widths, DSP parallelization, and other mutable parameters as top-level C directives. We've already seen some success in this approach; we were able to quickly re-target our original ZCU111-based design to the cost-effective, RFSoC4x2 platform shortly after its release. Going forward, these techniques are expected to continue helping us quickly leverage advances in the RFSoC product line. 

\subsection{UVOIR MKID Resolving Power}
Resolving power is a critical UVOIR MKID capability. As discussed in Sec. \ref{sec:rp}, the measured resolving power is a conglomeration of many factors and it can be challenging to identify performance requirements that can meaningfully inform design trade-offs. Several recent superconducting detector digital readouts have derived noise requirements from the cryogenic signal chain\cite{fruitwala_second_2020, yu_slac_2023}. \citet{fruitwala_second_2020} showed the Gen2 UVOIR MKID readout system IF loopback phase noise floor was around -90 dBc/Hz and was lower than the expected phase noise floor contributed by the first-stage, low-noise cryogenic HEMT amplifier ($\simeq$ -85 dBc/Hz)\cite{fruitwala_second_2020}. Here, the measured phase noise floor in the Gen3 system is 5-10 dB higher than this calculated HEMT noise floor (see Fig. \ref{fig:single_channel_psd} and Fig. \ref{fig:psd_floors}). This is not surprising given Gen3 runs twice as many tones in twice the bandwidth as compared to Gen2. We showed increasing the number of tones raises the system noise in Sec. \ref{sec:multi_tone_deg} and the larger bandwidth is expected to harbor more intermoduation products, reflections, and other spurious signal distortion.

The HEMT phase noise floor is not even the relevant cryogenic amplifier limit in this work because we used a quantum-noise-limited TWPA for the first-stage. Integrating a TWPA before the HEMT lowers the effective cryogenic amplifier noise temperature from around 5 K to close to 0.5 K\cite{zobrist_wide-band_2019}, further increasing the gulf between the expected phase noise contributions of the cryogenic signal chain and the Gen3 readout electronics. Despite having higher phase noise than the cryogenic amplifiers, Gen3 achieves respectable resolving power--competitive with dedicated analog electronics in the case of one tone\cite{zobrist_wide-band_2019} and comparable to Gen2 in the case of 2048 tones\cite{walter_mec_2018}.

Our results support the idea that the resolving power is predominately limited by the detector itself. \citet{zobrist_wide-band_2019} showed the addition of the first stage TWPA in the analog readout system did not improve resolving power as much as the reduction in phase noise would have predicted, indicating a detector limitation. We see a similar pattern in this work where the phase noise in Gen3 is higher than what was achieved in Gen2, yet the resolving power is comparable. 

This detector resolving power limitation has been attributed to a variable amount of the incident photon energy escaping into the non-photosensitive detector substrate, leading to a variable detector response. This problem, known in some literature references as ``hot phonon escape"\cite{kozorezov_quasiparticle-phonon_2000, kozorezov_electron_2007, kozorezov_phonon_2008, de_visser_phonon-trapping-enhanced_2021, zobrist_membraneless_2022}, is not captured in the phase noise power spectral density and is not improved by reduced phase noise in cryogenic or room temperature electronics. \citet{zobrist_membraneless_2022} fabricated a UVOIR MKID using a bilayer process that reflects energy back into the photosensitive region, recovering $R=20$ at 814 nm. Unfortunately, this bilayer process has not been demonstrated with large, array-style MKID devices and may be impracticable due to difficulties with indium processes. Solving hot phonon escape is an active area of UVOIR MKID design and fabrication research, but for now it presents a fundamental limitation in resolving power.

Our results coupled with previous work suggests UVOIR MKID readout systems should strive to achieve detector-limited resolving power as opposed to holding the phase noise under the contributions of the cryogenic amplifiers. With the addition of quantum-noise-limited TWPAs, this requirement becomes challenging and may come at the cost of increased system complexity and decreased scalability only to have the resolving power limited by the detectors in the end. In this work, we achieve detector-limited resolving power in the best-case scenario with one tone but we see degradation when running all 2048 tones through the readout chain (see Fig. \ref{fig:multi_tone}). The best-case scenario performance is useful as a tool in lab to characterize devices and inform design and fabrication updates; however, going forward we will strive to achieve detector limited resolving power at scale. Such a solution may require direct RF and or integration of more complex signal processing techniques such as tone tracking\cite{yu_slac_2023}, a resonator-based coordinate transform\cite{zobrist_improving_2021}, and a longer matched filter (Fig. \ref{fig:optimal_filter_expl}).

\section{Conclusion}
In conclusion, we have built a new MKID digital readout based on the RFSoC platform. The system fits twice as many MKID pixels onto a board that can be run with a fifth of the power of the previous system. Gen3 is also less than half the size and a tenth the weight of the Gen2 system. Gen3 utilizes high-level tools which simplify interacting with and programming the FPGA and make it easier for scientists to use, maintain, and upgrade. The system was demonstrated in lab and achieved high-fidelity, detector-limited resolving power in the case of few resonators and showed minimal degradation in $R$ with more tones. Experiments showed the degradation is linked to intermodulation products, image tones, and spurs in the analog domain which may be improved with future iterations of the IF board or a switch to a direct RF approach. 

Presently, we are continuing to update and improve the digital domain. We plan to adjust the rounding mode to an unbiased algorithm and include intelligent scaling to maximize dynamic range in the programmable logic. We are also working on new algorithms and techniques to expand the taps in the matched filter for improved noise rejection. Overall, we believe this readout system provides a maintainable, scaleable platform to work towards megapixel MKID arrays and future space-based deployment.

\begin{acknowledgments}
We would like to thank our collaborators at Fermilab: Gustavo Cancelo, Leandro Stefanazzi, and Ken Treptow for helping us get started with the RFSoC and for their valuable consultations on digital signal processing approaches.

We would also like to thank Ross Martin at Bit by Bit Signal processing for sharing his expertise on high-performance FFTs and for providing us with a power-efficient FFT that will enhance future design flexibility. 

Lastly, we would like to thank Jack Hickish, Mitch Burnett, and Dan Werthimer along with the rest of the CASPER collaboration for their early advice on FPGA-based signal processing approaches and for providing insight on RFSoC clocking and overall performance. 

\end{acknowledgments}

\bibliography{final_bib}

\end{document}